\documentclass[superscriptaddress,aps,prl,reprint,longbibliography]{revtex4-1}

\usepackage{amssymb,amsfonts,amsmath}
\usepackage{graphicx} % Include figure files
\usepackage{bm}        % bold math
\usepackage{color}
\usepackage{float}
\usepackage{subfigure}

\newcommand{\Ca}[0]{$^{40}$Ca$^+$}
\newcommand{\tP}[0]{\tau_{\textrm{P}_{3/2}}}
\newcommand{\AS}[0]{\textrm{A}_{\textrm{P}_{3/2}\rightarrow\textrm{S}_{1/2}}}
\newcommand{\ADf}[0]{\textrm{A}_{\textrm{P}_{3/2}\rightarrow\textrm{D}_{5/2}}}
\newcommand{\ADt}[0]{\textrm{A}_{\textrm{P}_{3/2}\rightarrow\textrm{D}_{3/2}}}
\newcommand{\DP}[0]{P$_{3/2}\leftarrow$D$_{5/2}$}
\newcommand{\SD}[0]{D$_{5/2}\leftarrow$S$_{1/2}$}

\newcommand{\DPS}[0]{D$_{3/2} \leftrightarrow$P$_{1/2}\leftrightarrow$S$_{1/2}$}
\newcommand{\tauP}[0]{6.639(42)} % our measured value - update here!
\newcommand{\tj}[6]{ \begin{pmatrix}
   #1 & #2 & #3 \\
   #4 & #5 & #6 
  \end{pmatrix}}
\newcommand{\D}[1]{\mathcal{D}_{\textrm{P}_{#1/2}\rightarrow\textrm{S}_{1/2}}}

\begin{document}

\title{Combining experiments and relativistic theory for establishing accurate radiative quantities in atoms: the lifetime of the $^2$P$_{3/2}$ state in \Ca{}}

\author{Ziv Meir}
\email{ziv.meir@unibas.ch}
\affiliation{Department of Chemistry, University of Basel, Klingelbergstrasse 80, Basel 4056, Switzerland.}
\author{Mudit Sinhal}
\affiliation{Department of Chemistry, University of Basel, Klingelbergstrasse 80, Basel 4056, Switzerland.}
\author{Marianna S. Safronova}
\affiliation{Department of Physics and Astronomy, 217 Sharp Lab, University of Delaware, Newark, Delaware 19716, USA.}
\affiliation{Joint Quantum Institute, NIST and the University of Maryland, College Park, Maryland, 20742}
\author{Stefan Willitsch}
\affiliation{Department of Chemistry, University of Basel, Klingelbergstrasse 80, Basel 4056, Switzerland.}
\date{\today}

\begin{abstract}
We report a precise determination of the lifetime of the (4p)$^2$P$_{3/2}$ state of \Ca{}, $\tP=\tauP$~ns, using a combination of measurements of the induced light shift and scattering rate on a single trapped ion. 
Good agreement with the result of a recent high-level theoretical calculation, $6.69(6)$~ns [Safronova \textit{et al.}, PRA \textbf{83}, 012503 (2011)], but a 6-$\sigma$ discrepancy with the most precise previous experimental value, $6.924(19)$~ns [Jin \textit{et al.}, PRL \textbf{70}, 3213 (1993)] is found.
To corroborate the consistency and accuracy of the new measurements, relativistically corrected ratios of reduced-dipole-matrix elements are used to directly compare our result with a recent result for the P$_{1/2}$ state, yielding a good agreement.
The application of the present method to precise determinations of radiative quantities of molecular systems is discussed.
\end{abstract}

\maketitle

% --------------------------------------------
%\subsection{Introduction} 
The knowledge of radiative lifetimes, transition rates, dipole-matrix elements and branching ratios in atoms and molecules is of great importance for, e.g.,  
experiments probing the electroweak force\cite{fortson93,roberts15,dutta16} in the search of physics beyond the standard model, 
for testing and improving atomic and molecular-structure theories\cite{safronova11,dutta16}, 
%for refining estimations of black-body-radiation shifts\cite{safronova11b} important for evaluating the accuracy of atomic clocks\cite{keller19} 
for the development of atomic clocks\cite{dutta16,safronova11b,keller19}
and for the interpretation of astronomical data\cite{carlsson12}. 

Traditionally, measurements of such quantities relied on atomic beams and short-pulse laser excitations. For instance, the last experimental evaluation of the radiative lifetime of the (4p)$^2$P$_{3/2}$ state of \Ca{} is dated 20 years back\cite{rosner97} and the most precise value was measured more than 25 years ago\cite{jin93} using such methods. Meanwhile, advances in experimental technology have enabled the control of single trapped atomic ions on the quantum level which led to the development of extremely precise atomic clocks\cite{huntemann16a,brewer19} and to a leading technology for quantum computers\cite{monz16a,linke17a}.

Here, we exploit the high-fidelity control achievable over a single trapped ion to establish a novel method to measure the complete set of lifetimes, transition rates and reduced-dipole-matrix elements of atomic transitions using the complementarity of dispersive and absorptive light-matter interactions and by combining precise experimental measurements with high-level theoretical calculations. To illustrate our approach, we determine the lifetime of the P$_{3/2}$ state in \Ca{} with high precision to $\tP=\tauP$~ns.

While our present value is in excellent agreement with a recent theoretical prediction using a high-precision relativistic all-order method\cite{safronova11} (6.69(6)~ns), it shows a 6-$\sigma$ fold discrepancy with the most precise previous value of Ref. \cite{jin93} (6.924(19)~ns). Interestingly, a similar discrepancy with these 25-years old results\cite{jin93} on the one hand and an agreement with the theoretical calculations\cite{safronova11} on the other hand was also established in a recent measurement of the lifetime of the (4p)~$^2$P$_{1/2}$ state of Ca$^+$ \cite{hettrich15}.

To corroborate the accuracy of the new measurements, highly precise theoretical ratios of reduced dipole matrix elements\cite{safronova11} were used to compare our result of the lifetime of the P$_{3/2}$ state with the recent results on the P$_{1/2}$ state \cite{hettrich15} with good agreement. Conversely, precise values of transition properties for a variety of states can be determined from their measurement for just a single state using the theoretical reduced-dipole-matrix elements. Elaborating on this combination of experiment and theory, previous measurements of the radiative branching ratios of the P$_{3/2}$ \cite{gerritsma08} and P$_{1/2}$ \cite{ramm13} states in Ca$^+$ were compared with excellent agreement and improved values of the polarizabilities of the (4s)$^2$S$_{1/2}$, (3d)$^2$D$_{3/2}$ and (3d)$^2$D$_{5/2}$ states of Ca$^+$ are recommended. The present approach for establishing values of radiative quantities can readily be generalized to non-atomic systems. In particular, it opens up new perspectives for precision measurements on molecules discussed at the end of this paper.

The use of a combination of absorptive and dispersive ion-light interactions to determine dipole-matrix elements and associated values was first demonstrated by Hettrich \textit{et al.} \cite{hettrich15}. In this work, a different variant of that technique which was proposed by Gerritsma \textit{et al.} \cite{gerritsma08} and was recently applied by Arnold \textit{et al.} \cite{arnold19} to measure the polarizability of Lu$^+$ was used.  

\begin{figure}
	\centering
	\includegraphics[width=\linewidth,trim={14.5cm 3.5cm 23cm 10.5cm},clip]{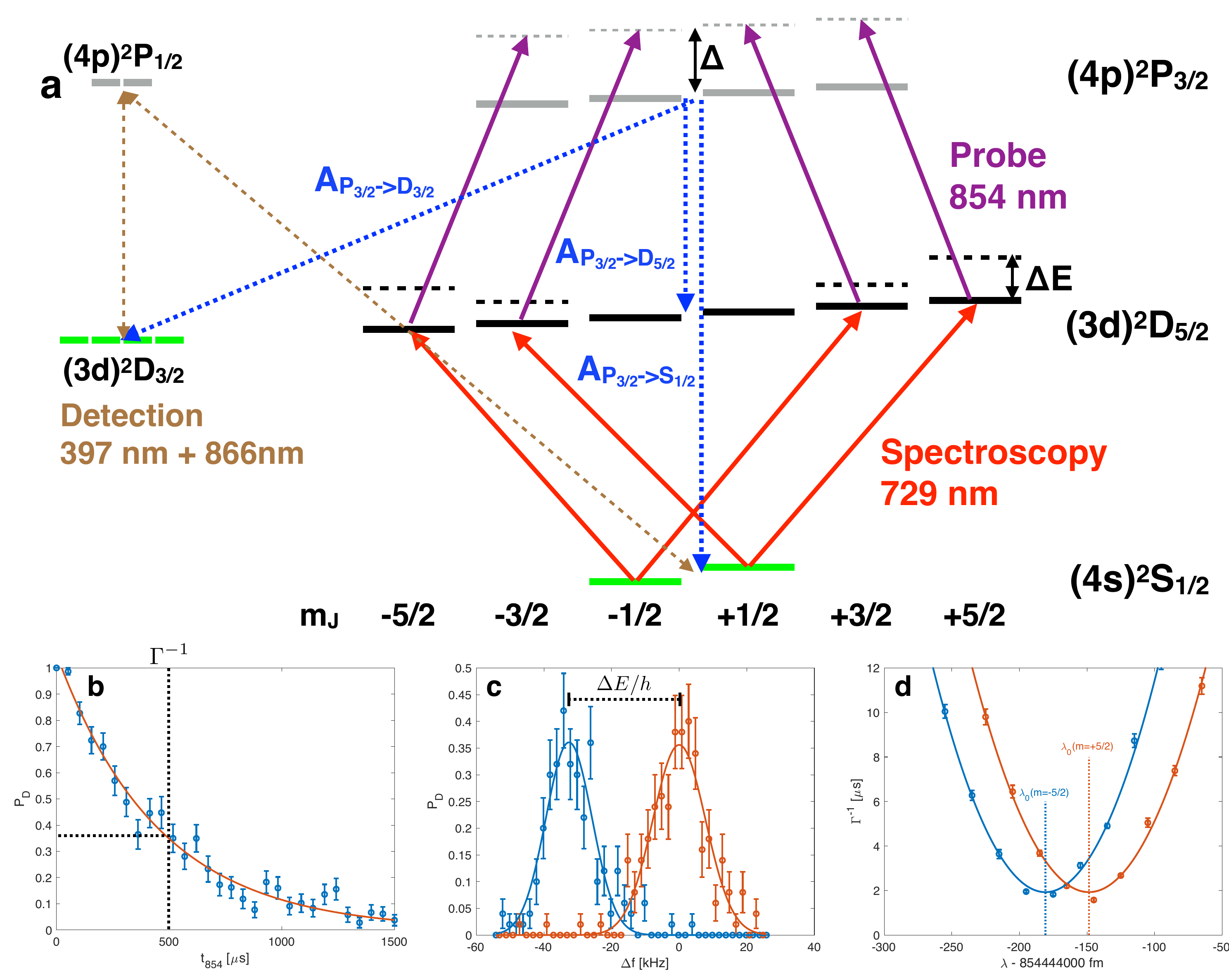}
	\caption{a) Energy diagram illustrating the present experimental scheme.
	b) Typical measurement instance of the scattering rate from the ``dark" state to ``bright" states. Data errors (blue) symbols are binomial projection-noise errors. The red line is a fit to an exponential function. Dashed lines are guides indicating the measured value of the scattering rate. 
	c) Typical measurement instance of the ac-Stark shift. Data errors (probe-beam on - blue; probe-beam off - red) are binomial projection noise. Lines are fits to Gaussian functions. The dashed line indicates the measured ac-Stark shift. 
	d) Typical measurement instance of the \DP{} resonance wavelength. The blue and red symbols are scattering rate measurements for ions prepared in the D$_{5/2}$(m=-5/2) and D$_{5/2}$(m=+5/2) states, respectively. Errors are 1-$\sigma$ confidence intervals of the exponential fit to data similar to the ones shown in panel (a). The solid lines are fits to a parabola. The dotted lines indicate the resonance wavelength of each Zeeman transition.}
	\label{fig:energy_diagram}
\end{figure}

% -----------------------------------------
%\subsection{Experiment}
Our measurement scheme is depicted in Fig. \ref{fig:energy_diagram}. A probe beam detuned from the \DP{} transition at 854 nm by $\Delta$ induces an ac-Stark shift, $\Delta E$, of magnitude (in Joules):
\begin{equation}\label{eq:acshift}
    \Delta E/h=\frac{1}{2\pi}\frac{\Omega^2}{4\Delta}.
\end{equation}
Here, $h$ is the Planck constant and $\Omega$ is the Rabi frequency. The probe beam also transfers population from the D$_{5/2}$ ``dark" state to the S$_{1/2}$ and D$_{3/2}$ ``bright" states by photon scattering via the P$_{3/2}$ state. The S$_{1/2}$ and D$_{3/2}$ states are considered ``bright" since both participate in the closed cycle fluorescence transition \DPS{}. The rate by which population is transferred is given by:
\begin{equation}\label{eq:scatter}
    \Gamma=\left(\AS+\ADt\right)\frac{\Omega^2}{4\Delta^2}.
\end{equation}
Here, $\AS$ and $\ADt$ are the transition rates (in s$^{-1}$) connecting the excited P$_{3/2}$ state with the ``bright" states.
The scattering rate, $\Gamma$ (in s$^{-1}$), also depends on the Rabi frequency which is difficult to determine with high accuracy in an experiment due to its dependence on the laser intensity and polarization. However, the ratio of the scattering rate and the light shift does not depend on the Rabi frequency which allows for a direct determination of the transition rates without the need for precise characterization of the probe beam intensity and polarization:
\begin{equation}\label{eq:ratio}
    \AS+\ADt=\frac{\Delta}{2\pi}\frac{\Gamma}{\Delta E/h}.
\end{equation}

Eqs. \ref{eq:acshift},\ref{eq:scatter} are approximations for the ac-Stark shift and scattering rate calculated from a second-order perturbation theory. The first approximation, $\Delta\gg\Omega$, neglects the line shape near resonance (e.g, the Lorentzian scattering profile). The second, $\Delta \ll \omega_0$, with $ \omega_0$ the transition's angular frequency, neglects co-rotating terms when performing the rotating-wave approximation. The third neglects contributions from transitions other than the \DP{}. The last neglects the finite lifetime of the D$_{5/2}$ state. All above approximations were treated as systematic shifts which are listed in Table \ref{tab:errors} and discussed in more details in the supplemental material (SM)\cite{sm}. For the chosen probe-beam detuning and intensity, these approximations hold to a high degree of accuracy compared to the measurement uncertainty and other systematic shifts such that they can be neglected. Further shifts and errors of the measurement will be discussed later in the text.

% -----------------------------------------
% Experiment

Our experimental apparatus consists of a linear Paul trap for trapping single Ca$^+$ ions at mK temperatures using Doppler cooling\cite{meir19a}. A narrow-linewidth laser on the \SD{} transition at 729 nm was used to prepare the ion in one of the meta-stable Zeeman states (m=$\pm$5/2,$\pm$3/2) of the D$_{5/2}$ electronic state and to perform precision spectroscopy on the \SD{} transition (see Fig. \ref{fig:energy_diagram}). A probe beam at 854 nm detuned from the \DP{} transition was used to induce scattering from and light shifts of the D$_{5/2}$ state. The probe beam was linearly polarized perpendicular to the external magnetic field such that it excited mostly $\Delta m=\pm1$ transitions. Detection beams at 397 nm and 866 nm which are in resonance with the \DPS{} cycling transitions were used to distinguish between ``bright" and ``dark" states. 

The scattering rate, $\Gamma$, was measured by recording the ``dark" population, P$_\textrm{D}$, as a function of the probe time, t$_{854}$ (see Fig. \ref{fig:energy_diagram}b). The ion was prepared in the D$_{5/2}$ state using a $\pi$-pulse of the spectroscopy laser followed by a projection pulse of the detection beams which enables post-selection of experiments starting in the D$_{5/2}$ state only. The probe beam was then turned on using an acousto-optic modulator (AOM) 5 $\mu$s before the experiment began in order to avoid any AOM latency (typically less than 1 $\mu$s). Experimental data was fitted with an exponential function, $\exp{(-\Gamma \left(t_{854}-t_0\right))}$, to extract the scattering rate. Here, $t_{854}$ is the experiment time and $t_0$ accounts for the fact that the AOM was turned on before the experiment began.  %Typical run takes xx sec and results with 4\% uncertaint

The ac-Stark shift, $\Delta E$, was measured by performing Rabi spectroscopy on the \SD{} transition using the narrow-linewidth spectroscopy laser (see Fig. \ref{fig:energy_diagram}c). The probe beam was switched on 5 $\mu$s before the spectroscopy pulse to avoid latency effects. The experimental cycles were interlaced with the probe beam on and off in order to cancel errors from decoherence mechanisms such as magnetic-field and spectroscopy-laser-phase fluctuations. The energy shift between the transitions with the probe beam on and off was determined by comparing the centers of Gaussian fits for each of the observed lines.

The probe-beam wavelength, $\lambda$, was monitored by and locked to a wavemeter (HighFinesse WS-U 30) and scanned by changing the locking set point. The center wavelength of the \DP{} transition, $\lambda_0$, was found by scanning the probe-beam wavelength across resonance using a weak probe-beam power below saturation intensity while measuring the scattering rate from ``dark'' to ``bright'' states (see Fig. \ref{fig:energy_diagram}d). The central wavelength of the transition was determined by fitting the inverse scattering rate to a second-order polynomial. The resonance frequency starting from both D$_{5/2}$(m=$\pm5/2$) Zeeman states was measured to account for Zeeman splittings in a magnetic field of 4.609(2) Gauss. The magnetic field was also measured with high precision on the \SD{} transition using the narrow spectroscopy laser. The probe-beam detuning, $\Delta(m)=2\pi c\left(1/\lambda-1/\lambda_0(m)\right)$, was determined for each of the Zeeman states. 

\begin{figure*}
	\centering
	\includegraphics[trim={0.5cm 11.5cm 6cm 13.5cm},clip=true,width=\textwidth]{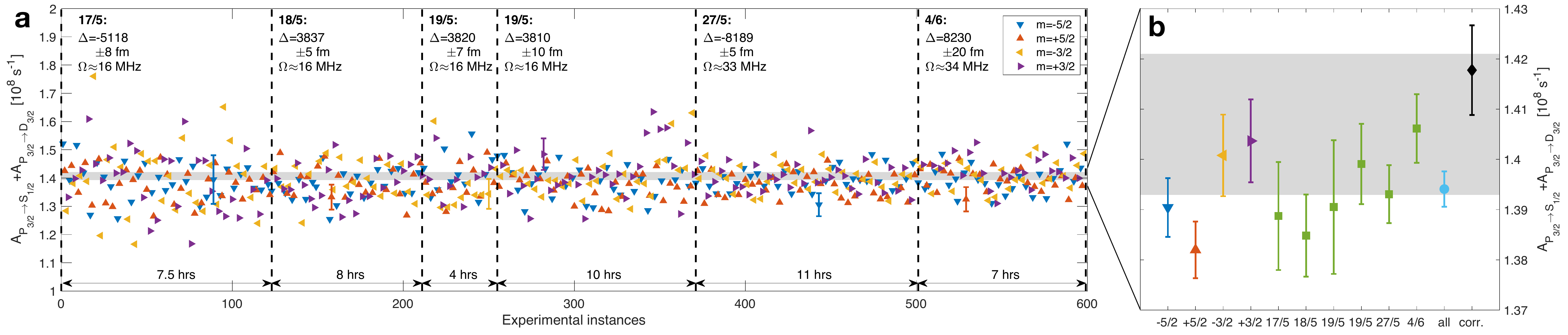}\\
	\caption{a) Measurements results of the transition rates $\AS+\ADt$ prior to systematic shifts corrections. Initial Zeeman states of the D$_{5/2}$ state are marked with different colors and symbols (see legend). Dashed lines indicate the re-measurement of the \DP{} resonance wavelength and the change of probe-beam detuning and intensity settings. Typical error bars representing 1-$\sigma$ confidence intervals of the fits of the scattering rate, ac-Stark shift and detuning measurements are given. The grey area represents the theoretical value and their uncertainty\cite{safronova11}. b) Averaged results of the different Zeeman states (triangles), laser settings (green squares) and all data (Light blue circle). All results up to this point are prior to systematic shifts corrections. The error bars are a combination of statistical standard errors and measurement fitting errors. For the result represented by the black diamond, systematic shifts and additional uncertainties were included (see Table \ref{tab:errors}).}
	\label{fig:results_time}
\end{figure*}

Eq. \ref{eq:ratio} was used to determine the transition rates for each experimental instance. Our measurements of the scattering rate and the ac-Stark shift were repeated 600 times interlacing between different initial Zeeman states, m=$\pm5/2,\pm3/2$, of the D$_{5/2}$ state. Every few hours, the probe-beam detuning and intensity were changed and the transition center wavelength was re-measured to reduce errors due to drifts in the probe-beam frequency. The scattering rate and ac-Stark shift measurements were then continued for a few more hours. In total, four different combinations of probe-beam detunings and intensities were measured for a duration of almost 50 hours (see Fig. \ref{fig:results_time}a).

% -----------------------------------------
% Results

All measurements were averaged to determine the sum of the transition rates $\AS+\ADt$ (see Fig. \ref{fig:results_time}b). The total uncertainty of our measurement (0.25\%) includes both the standard error of all individual measurements and the measurement errors arising from the confidence intervals of the fits (see Table \ref{tab:errors}). The measurements were also averaged separately for each Zeeman state and each different detuning and intensity setting. %All results agree within the uncertainty limits indicating no systematic shifts between the chosen Zeeman levels or the probe-beam detunings and intensities within the precision of our measurements. 

Possible systematic shifts for this type of measurement are listed in Table \ref{tab:errors} (see SM for detailed discussion and derivation\cite{sm}). The most dominant one is the effect of inelastic Raman scattering\cite{ozeri05} that changes the Zeeman state in the D$_{5/2}$ manifold before scattering to the ``bright'' states. This event changes the Rabi frequency during the scattering-rate measurement and thus shifts the measured value of the scattering rate (see Eq. \ref{eq:scatter}). On the other hand, inelastic Raman scattering events will not shift the value of the ac-Stark shift due to the Zeeman selectivity of the narrow spectroscopy laser.

To evaluate this shift, a numerical calculation of the dynamical optical Bloch equations (DOBE) describing our system was performed (see e.g. \cite{meir17b,meir18a}). The ``dark'' population decay was determined for different initial Zeeman states of the D manifold and was found to deviate from a single exponential decay, as expected due to the small leak into different Zeeman states of the D manifold. Instead, a sum of three exponents was used to better describe the decay owing to the three different Rabi coupling in the D-manifold\cite{sm}. From a fit of all the scattering data the inelastic Raman scattering shift is extracted (see Table \ref{tab:errors}). Notably, while the systematic shift is larger than our measurement uncertainty, it is almost the same for the $\pm5/2,\pm3/2$ Zeeman states. The Raman inelastic scattering effect was experimentally verified by interlacing measurements between $\pm5/2$ states to $\pm1/2$ states which features opposite and distinctively measurable systematic shifts\cite{sm}.

Even after accounting for the inelastic Raman scattering systematic shift, a discrepancy of 2-$\sigma$ between the $\pm$3/2 and the $\pm$5/2 measurements still remains. Since this discrepancy cannot be accounted for, it is added as an uncertainty of 0.56\% which is the dominant contribution to the error of this measurement.

The second-most dominant shift is due to thermal effects in the probe-beam AOM. While the rise time of the AOM is less than 1 $\mu$s, it takes about 15 $\mu$s (1/e) for the AOM to reach stable operation. For the ac-stark shift measurements, due to a 2 ms D-state repump pulse just before the measurement starts, the AOM is in steady-state operation and no systematic shifts were observed experimentally\cite{sm}. However, for the scattering-rate measurements, there is almost a ms where the probe beam is turned off before the measurement starts. We experimentally verified and quantified this systematic shift by omitting the first data points of the scattering from the analysis\cite{sm}. 

\begin{table}
  \begin{center}
    %\begin{tabular}{l|c|r}
%    \begin{tabular*}{\linewidth}{l@{\extracolsep{\fill}}lcr}
    \begin{tabular}{l r r}
      \hline
      \hline
      Effect \hspace{3.8 cm} & Shift [\%] & Uncertainty [\%] \\
      \hline
      Statistical standard Error & ... & $\pm$0.20\\
      Fit error & ... & $\pm$0.16\\
      \hspace{0.5 cm} Total statistical error & ... & $\pm$0.25\\
      Line shape & $<+0.008$ & ... \\
      Rotating-wave approximation & +0.00005 & ...\\
      Other lines & -0.0003 & ...\\
      D$_{5/2}$ state lifetime & -0.04 & ... \\
      Detection threshold & ... & $\pm$0.04\\
      Finite detection time & -0.06 & ... \\
      AOM thermal effect & +0.41 & $\pm$0.14 \\
      Motion-induced Doppler shifts & $<-0.0001$ & ... \\
      Inelastic scattering ($m_D=\pm5/2$) & +1.29 & ... \\
      Inelastic scattering ($m_D=\pm3/2$) & +1.46 & ... \\
      Off-resonant Raman coupling & $<-0.001$ & ... \\
      Zeeman states discrepancy & ... & $\pm$0.56 \\
      \hspace{0.5 cm} \textbf{Total shifts \& errors} & \textbf{+1.70} & \textbf{$\pm$0.63} \\
      \hline
      \hline
    \end{tabular}
    \caption{Systematic shifts and experimental uncertainties. The symbol $<$ is used to indicate that the calculated absolute value of the shift is an upper bound. For shifts with a $+(-)$ sign, the measured value should be increased (decreased) accordingly.}
    \label{tab:errors}
  \end{center}
\end{table}

In Fig. \ref{fig:results_time}b, the measured value of the summed transition rates $\AS+\ADt$ corrected for all systematic effect is shown and compared to the non-corrected value. Our result of $\AS+\ADt=1.4178(89)\cdot10^8$ s$^{-1}$ agrees well with a theoretical calculation ($1.407(14)\cdot10^8$ s$^{-1}$)\cite{safronova11}.

\begin{figure}
	\centering
	\includegraphics[width=\linewidth,trim={3cm 0cm 2.5cm 1cm},clip]{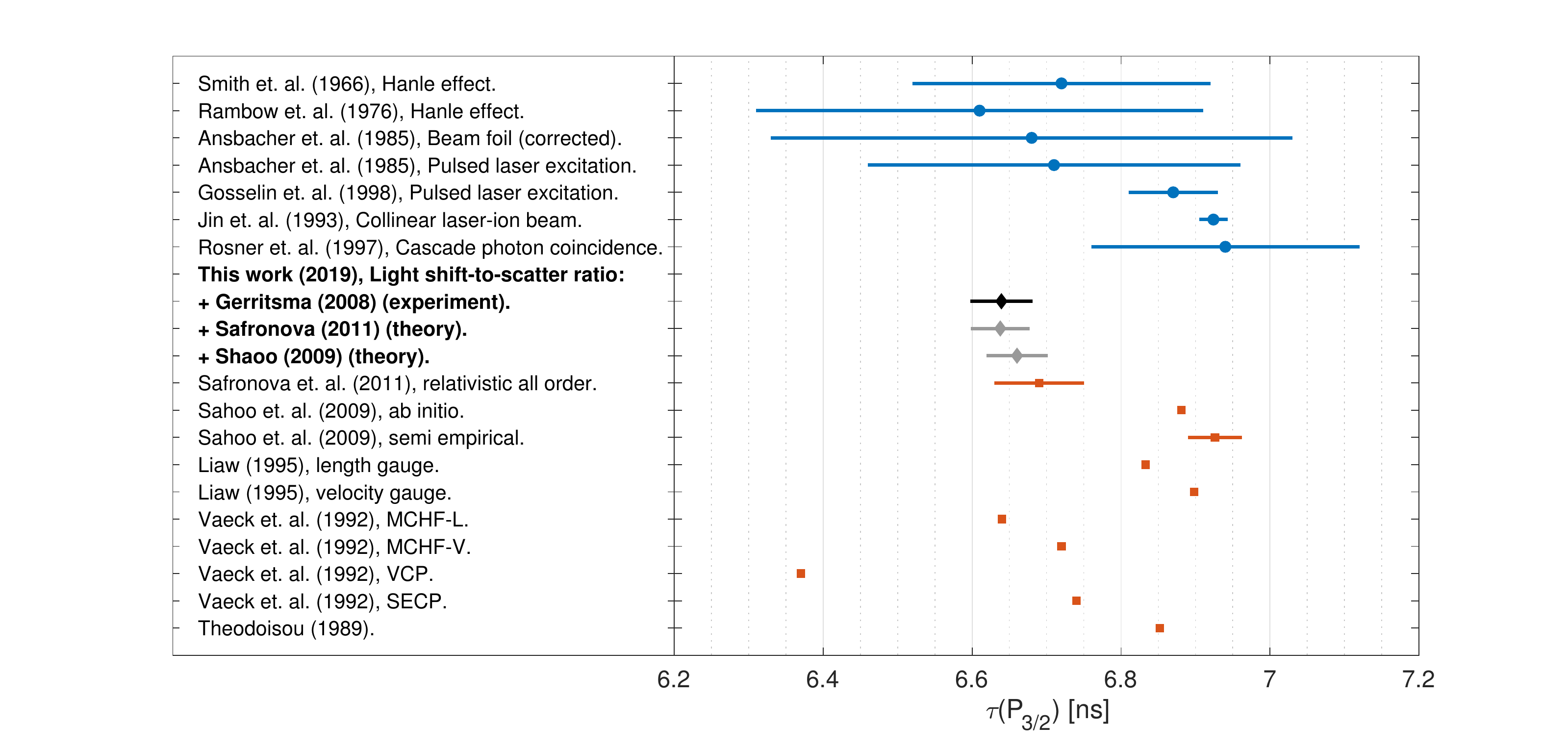}
	\caption{Comparison of different experimental (blue circles) \cite{smith66,rambow76,ansbacher85,gosselin88,jin93,rosner97} and theoretical (red squares) \cite{theodosiou89,vaeck92,liaw95,sahoo09,safronova11} values for the P$_{3/2}$ lifetime in \Ca{} to this work (black diamond). Some of the theoretical works (red squares with no error bars) did not quote errors. For the recommended lifetime value of this work (black diamond), an experimental branching value from \cite{gerritsma08} was used. Calculation of the lifetime from the $\ADf$ theoretical value of \cite{safronova11} and \cite{sahoo09} is also given (grey diamonds).}
	\label{fig:comparison}
\end{figure}

The lifetime of an excited atomic state is given by the inverse of the sum of the transition rates from that excited state. For the P$_{3/2}$ state in Ca$^+$ one gets
\begin{equation}\label{eq:lifetime}
    \tP=\frac{1}{\AS+\ADt+\ADf}.
\end{equation}
Our measurements determined the sum $\AS+\ADt$. The value of $\ADf$ contributing to the P$_{3/2}$ state lifetime can be measured using our technique by switching to a different probe beam that connects the S$_{1/2}$ and P$_{3/2}$ states. Here, however, a high-precision experimental value for the branching ratio, $\textrm{R}_{\textrm{P}_{3/2}\rightarrow\textrm{D}_{5/2}}=0.0587(2)$ \cite{gerritsma08} was used to determine the recommended value for the total lifetime, $\tP=(1-\textrm{R}_{\textrm{P}_{3/2}\rightarrow\textrm{D}_{5/2}})/(\AS+\ADt)=\tauP$ ns. In addition, two different theoretical values for the value of $\ADf$ \cite{safronova11,sahoo09} with their respective uncertainties were used to verify our experimental value for the lifetime. Since the value of $\ADf$ is one order of magnitude smaller than $\AS$, even though the two theories disagree within a few standard deviations, all calculated lifetime values agree within the uncertainty limits (see Fig. \ref{fig:comparison}).

The branching ratio, $\textrm{R}_{\textrm{P}_{3/2}\rightarrow\textrm{S}_{1/2}}=0.9347(3)$ \cite{gerritsma08}, is further used to calculate the transition rate, $\AS=\textrm{R}_{\textrm{P}_{3/2}\rightarrow\textrm{S}_{1/2}}/\tP$, and the reduced dipole matrix element,
\begin{equation}\label{eq:matrix}
    \D{3}^2=\left(2J_{\textrm{P}_{3/2}}+1\right)\AS\frac{3\epsilon_0\hbar}{8\pi^2}\lambda_{\textrm{P}_{3/2}\rightarrow\textrm{S}_{1/2}}^3.
\end{equation}
The value of $\D{3}=4.115(13)$ ea$_0$ is compared to the value of $\D{1}=2.8928(43)$ ea$_0$ (Hettrich \textit{et al.} \cite{hettrich15}) using a high precision theoretical ratio $\D{3}/\D{1}=1.4145(1)$ \cite{safronova11} yielding 4.092(6) ea$_0$. This way, both the experimental values are directly compared without loss of uncertainty and agree to within 1.6 $\sigma$.

\mbox{The reduced-dipole-matrix-elements ratios,} $\mathcal{D}_{\textrm{P}_{3/2}\rightarrow\textrm{D}_{5/2}}/\mathcal{D}_{\textrm{P}_{3/2}\rightarrow\textrm{D}_{3/2}}$=3.0068(13), \mbox{$\mathcal{D}_{\textrm{P}_{3/2}\rightarrow\textrm{D}_{5/2}}/\mathcal{D}_{\textrm{P}_{1/2}\rightarrow\textrm{D}_{3/2}}$=1.3421(4) and} $\mathcal{D}_{\textrm{P}_{3/2}\rightarrow\textrm{D}_{3/2}}/\mathcal{D}_{\textrm{P}_{1/2}\rightarrow\textrm{D}_{3/2}}$=0.44634(6), are further used to compare the experimental branching ratios of the P$_{3/2}$ with those of the P$_{1/2}$ measured by Ramm \textit{et al.} \cite{ramm13} with excellent agreement. The converted P$_{1/2}$ values are of better precision than the directly measured P$_{3/2}$ ones (Table \ref{tab:comp}).
The matrix-element ratios are of such high precision due to common electronic-correlations contributions for transitions involving different fine-structure components. 

\begin{table}
  \begin{center}
   \resizebox{\linewidth}{!}{
   \begin{tabular}{|l|c|c|c|}
      \hline
      \hline
       & P$_{1/2}$ exp. & P$_{3/2}$ & P$_{3/2}$ \\
       & converted &  experiment & theory \\
      \hline
      $\textrm{R}_{\textrm{P}_{3/2}\rightarrow\textrm{S}_{1/2}}$ & 0.93463(9) \cite{ramm13,safronova11} & 0.9347(3) \cite{gerritsma08} & 0.9340(9) \cite{safronova11} \\
      $\textrm{R}_{\textrm{P}_{3/2}\rightarrow\textrm{D}_{5/2}}$ & 0.05876(8) \cite{ramm13,safronova11} & 0.0587(2) \cite{gerritsma08} & 0.0593(8) \cite{safronova11} \\
      $\textrm{R}_{\textrm{P}_{3/2}\rightarrow\textrm{D}_{3/2}}$ & 0.006602(7) \cite{ramm13,safronova11} & 0.00661(4) \cite{gerritsma08} & 0.00667(9) \cite{safronova11} \\
      \hline
      $\mathcal{D}_{\textrm{P}_{3/2}\rightarrow\textrm{S}_{1/2}}$ & 4.092(6) \cite{hettrich15} & 4.115(13) \cite{tw,gerritsma08} & 4.099(18) \cite{safronova11} \\
      $\mathcal{D}_{\textrm{P}_{3/2}\rightarrow\textrm{D}_{5/2}}$ & 3.283(6) \cite{hettrich15,ramm13} & 3.300(12) \cite{tw,gerritsma08} & 3.306(18) \cite{safronova11} \\
      $\mathcal{D}_{\textrm{P}_{3/2}\rightarrow\textrm{D}_{3/2}}$ & 1.092(2) \cite{hettrich15,ramm13} & 1.097(5) \cite{tw,gerritsma08} & 1.100(6) \cite{safronova11} \\
      \hline
      \hline
    \end{tabular}}
    \caption{Translation of P$_{1/2}$ experimental values of reduced-dipole-matrix elements\cite{hettrich15} and branching ratios\cite{ramm13} to P$_{3/2}$ values and their comparison to the measured experimental values of this work and Ref. \cite{gerritsma08} and the theoretical values of Ref. \cite{safronova11}. The translation is done using high-precision theoretical ratios of reduced-dipole-matrix elements\cite{safronova11}.}
    \label{tab:comp}
  \end{center}
\end{table}

The directly measured values of reduced-dipole-matrix elements and a high-precision measurement of the differential polarizability $\alpha_0(3d_{5/2})-\alpha(4s)=-44.079(13)$ a.u. by Huang \textit{et al.} \cite{huang19} are used to extract improved recommended values for the scalar polarizabilities $\alpha(4s)=76.40(32)$, $\alpha_0(3d_{3/2})=31.72(22)$ and $\alpha_0(3d_{5/2})=32.32(32)$ and tensor polarizabilities $\alpha_2(3d_{3/2})=-17.18(8)$ and $\alpha_2(3d_{5/2})=-24.42(17)$. All values are in atomic units (see \cite{sm} for further details).

% molecular ions stuff
A particularly attractive application of the present method is the measurement of the lifetimes of quantum states of molecular ions within the framework of a quantum-logic experiment\cite{wolf16a,chou17a,meir19a}. Consider, e.g, the N$_2^+$ molecular ion in its electronic (X$^2\Sigma_g^+$) and vibrational (v$''$=0) ground state\cite{tong10a, germann14a}. A probe beam consisting of a 1D optical lattice modulated at the trap frequency and detuned closely to an excited state such as the A$^2\Pi_u^+$ (v$'$=2) will induce an optical-dipole force proportional to the ac-Stark shift experienced by the molecule \cite{meir19a}. The force can be detected by a co-trapped atomic ion using quantum logic protocols\cite{koelemeij07a,hume11a,meir19a}. Upon scattering, the molecule will decay to a vibrational level of the X$^2\Sigma_g^+$ state according to Franck-Condon factors. Scattering into vibrational states other than the ground state (v$''$=0) will diminish the optical-dipole force due to the increased detuning, thus signalling the time of scattering. The ratio of the scattering rate and the ac-Stark shift gives the sum of all transitions rates $\sum_{\textrm{v}''\neq0}\textrm{A}_{2 \rightarrow \textrm{v}''}$ except one, A$_{2\rightarrow0}$, which can be extracted from the ac-Stark shift measurement. The inverse of the sum of all vibronic transitions rates gives the vibronic lifetime, $\tau_{\textrm{v}'=2}=1/\sum_{\textrm{v}''}{\textrm{A}_{2 \rightarrow \textrm{v}''}}$. This discussion only includes vibronic states. Rotational, fine and hyperfine structure can be considered in a similar fashion.

To summarize, measurements of transition rates and branching ratios were combined together with relativistic theory to achieve precision and to validate the accuracy of the lifetime of the (4p)$^2$P$_{3/2}$ excited state of Ca$^+$, $\tP=\tauP$ ns.
The present method can be used to measure transition rates and lifetimes in many types of ionic, atomic and molecular systems both for single and ensembles of particles.

We thank Anna Ladina Leder and Heinz Krummenacher who helped to perform the first proof-of-principle experiments during an undergraduate laboratory project. We also thank Roee Ozeri for helpful comments on the manuscript. This work was supported by the Swiss National Science Foundation, grant nr. CRSII5\_183579, and through the National Competence Centre in Research, Quantum Science and Technology (NCCR-QSIT). MS work was supported in part by U.S. NSF Grant No. PHY-1620687.

\bibliographystyle{apsrev4-1}
%\bibliography{Main-Feb19}

\newpage
\onecolumngrid

\section*{Supplemental Material}

\section{Systematic shifts}
According to Eq. [3] of the main text, the ratio of three experimentally measured parameters ($\Gamma$, $\Delta E$, $\Delta$) equals to the sum of the transitions rates $\AS+\ADf$ which we will denote as A from now on for brevity. This equality holds for the approximate equations of the ac-Stark shift (Eq. [1] of the main text) and the scattering rate (Eq. [2] of the main text). The experimentally measured values, however, follow the exact formulas for these parameters and hence the ratio given in Eq. [3] in the main text does not equate exactly to A, but it differs by a small amount,
\begin{equation*}
    x_i\equiv\frac{\Delta_i}{2\pi}\frac{\Gamma_i}{\Delta E_i/h}=\textrm{A}_i\left(1-\varepsilon_i\right).    
\end{equation*}
Here, $x_i$ is the value calculated from the measurements in the experimental instance $i$ and $\varepsilon_i$ is the systematic shift of that experimental instance. Positive $\varepsilon_i>0$ means that our measured value of $x_i$ should be increased by $\varepsilon_i$ since $\textrm{A}_i\approx x_i(1+\varepsilon_i)$ for $|\varepsilon_i|\ll1$.

In our experiment, we performed measurements with different laser powers and detunings while interlacing between different Zeeman states. For each instance of the experiment, $i$, we calculate a systematic shift, $\varepsilon_i$. Our best estimate for A is given by the mean of all our measurements:
\begin{equation*}
    \textrm{A}=\langle x_i\left(1+\varepsilon_i\right)\rangle=\langle x_i\rangle+\langle x_i\varepsilon_i\rangle\equiv\langle x_i\rangle\left(1+\varepsilon\right).
\end{equation*}
Here, $\langle x_i\rangle$ is the mean of all measured values of the transition rates without correction and
\begin{equation*}
    \varepsilon=\frac{\langle x_i\varepsilon_i\rangle}{\langle x_i\rangle}.
\end{equation*}
is the weighted mean of the systematic shifts of each experimental instance. The value of $\varepsilon$ calculated for different types of systematic shifts is given in The Table I of the main text.

\subsection{Line shape} 
The solution of a two-level system interacting with the classical electric field of an electromagnetic wave in the rotating-wave approximation gives rise to the well-known Lorentzian profile for the excited-state population \cite{kimble76}:
\begin{equation*}
    p_e=\frac{\Omega^2/4}{\Omega^2/2+\Delta^2+(1/\tP)^2/4}.
\end{equation*}
The scattering rate to ``bright" states which decouple from the two-level system is given by:
\begin{equation*}
    \Gamma=p_e \textrm{A}.%\cdot\left(\AS+\ADt\right).
\end{equation*}
For large enough detuning, $\Delta\gg\Omega,\ADf$, these equations approximate to Eq. [2] of the main text and give rise to a systematic shift:
\begin{equation*}
    \varepsilon_i\approx\frac{\Omega_i^2/2+(1/\tP)^2/4}{\Delta_i^2}.
\end{equation*}
We determine the Rabi frequency using Eq. [1] of the main text. We take the value of $\tP$ from Ref. \cite{safronova11}, $\tP=6.69$ ns. The mean systematic shift of all experimental instances is $\varepsilon < 7.7\cdot10^{-5}$ which is negligible compared to the measurement uncertainty. This shift is an upper bound since as the exact scattering rate decreases as compared to the approximated value (Eq. [2] of the main text) when approaching the resonance, the exact ac-Stark shift also decreases as compared to the approximated value (Eq. [1] of the main text). These two effects effectively cancel leading to a much smaller shift. Nevertheless, the upper bound is small enough such that it is not necessary to account for this effect in the present case.

\subsection{Rotating-wave approximation} 
Outside the rotating-wave approximation, the ac-Stark shift takes the form \cite{grimm00a,zhou10}:
\begin{equation*}
    \Delta E/h=\frac{1}{2\pi}\frac{\Omega^2}{4}\cdot\left(\frac{1}{\omega-\omega_0}-\frac{1}{\omega+\omega_0}\right).
\end{equation*}
Here, $\omega$ is the laser frequency and $\omega_0$ is the transition frequency such that: $\Delta=\omega-\omega_0$. The scattering rate outside the rotating-wave approximation is given by \cite{grimm00a},
\begin{equation*}
    \Gamma=\textrm{A}\frac{\Omega^2}{4}\left(\frac{\omega}{\omega_0}\right)^3\left(\frac{1}{\omega_0-\omega}+\frac{1}{\omega_0+\omega}\right)^2.
\end{equation*}
For $\Delta\ll\omega_0$ both equations approximate to Eq. [1] and Eq. [2] of the main text and give rise to a systematic shift:
\begin{equation*}
    \varepsilon_i=\Delta_i\left(\frac{3}{\omega_{0,i}}+\frac{1}{\omega_{0,i}+\omega}\right).
\end{equation*}
Note that this systematic shift depends on the sign of the detuning, and in our experiment we used both red and blue detuned probe lasers such that the systematic shifts partially cancel giving rise to $\varepsilon=5.1\cdot10^{-7}$. Nevertheless, the maximum absolute value of this systematic shift is $|\varepsilon_i|<3.4\cdot10^{-5}$ which is negligible with respect to our measurement uncertainty.

\subsection{Other lines}
The probe beam mainly interacts with the \DP{} transition near 854.4 nm and shifts both the D$_{5/2}$ and the P$_{3/2}$ levels. We monitored this ac-Stark shift by performing precision spectroscopy on the \SD{} transition using a narrow-linewidth laser beam at 729 nm as discussed in the main text. The probe beam interacts with all other allowed transitions from both the S$_{1/2}$ and the D$_{5/2}$ states. These interactions induce a systematic shift of the measured ac-Stark shift value. Due to the $\Delta^{-2}$ dependence of the scattering rate and the large detuning for any other transition, the scattering effect is negligible.

The dominant interaction of the probe beam other than with the \DP{} transition is with the P$_{1/2}\leftarrow$S$_{1/2}$ and the P$_{3/2}\leftarrow$S$_{1/2}$ transitions. The probe beam at 854.4 nm was highly red detuned from these transitions at 397 nm and 393 nm respectively. The S$_{1/2}$ level was shifted by -1 Hz to -3 Hz depending on the laser parameters. Our measured value of the ac-Stark shift is then composed of two contributions:
\begin{equation*}
    \Delta E=\Delta E_\textrm{PD}-\Delta E_\textrm{PS}.
\end{equation*}
The systematic shift of every experimental instance is given by,
\begin{equation*}
    \varepsilon_i=-\frac{\Delta E_{\textrm{PS},i}}{\Delta E_{\textrm{PD},i}}.
\end{equation*}
Note that, as in the case of the shift due to the rotating-wave approximation, we have cancellation of systematic shifts from blue and red detuned experiments. When we change the detuning from red to blue in the experiment, $\Delta E_\textrm{PD}$ either assumes positive or negative values while $\Delta E_\textrm{PS}$ is always negative resulting in $\varepsilon=-3.4\cdot10^{-6}$. Nevertheless, the maximal value of the systematic shift is $|\varepsilon_i|<6.5\cdot10^{-5}$ which is negligible compared to our experimental uncertainty. 

\subsection{Finite D$_{5/2}$ state lifetime}
Due to finite lifetime of the D$_{5/2}$ state, Eq. [2] of the main text is changed to:
\begin{equation*}
    \Gamma=\textrm{A}\frac{\Omega^2}{4\Delta^2}+\left(\textrm{A}_{\textrm{D}_{5/2}\rightarrow\textrm{S}_{1/2}}+\textrm{A}_{\textrm{D}_{5/2}\rightarrow\textrm{D}_{3/2}}\right).
\end{equation*}
Here, $\left(\textrm{A}_{\textrm{D}_{5/2}\rightarrow\textrm{S}_{1/2}}+\textrm{A}_{\textrm{D}_{5/2}\rightarrow\textrm{D}_{3/2}}\right)=\tau_{\textrm{D}_{5/2}}^{-1}$ are the two transition rates connecting the D$_{5/2}$ ``dark'' state to the S$_{1/2}$ and D$_{3/2}$ ``bright'' states which give rise to a finite lifetime, $\tau_{\textrm{D}_{5/2}}=1.1649(44)$ s \cite{shao17}, of this state. We experimentally verified this lifetime (with less precision) in our experiment to overrule spurious optical pumping effects. 

The systematic shift for each experimental instance is given by,
\begin{equation*}
    \varepsilon_i=-\frac{\tau_{\textrm{D}_{5/2}}^{-1}}{\Gamma_i},
\end{equation*}
and the mean systematic shift is $\varepsilon=-3.7\cdot10^{-4}$ which is small compared to our measurement uncertainty. 

\subsection{Detection threshold}
\begin{figure}
     \begin{center}
         \subfigure{\includegraphics[scale=0.23]{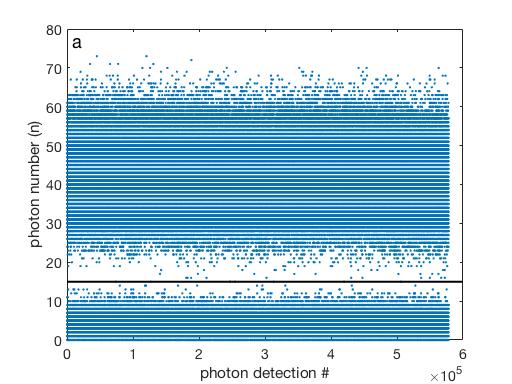}}
         \subfigure{\includegraphics[scale=0.23]{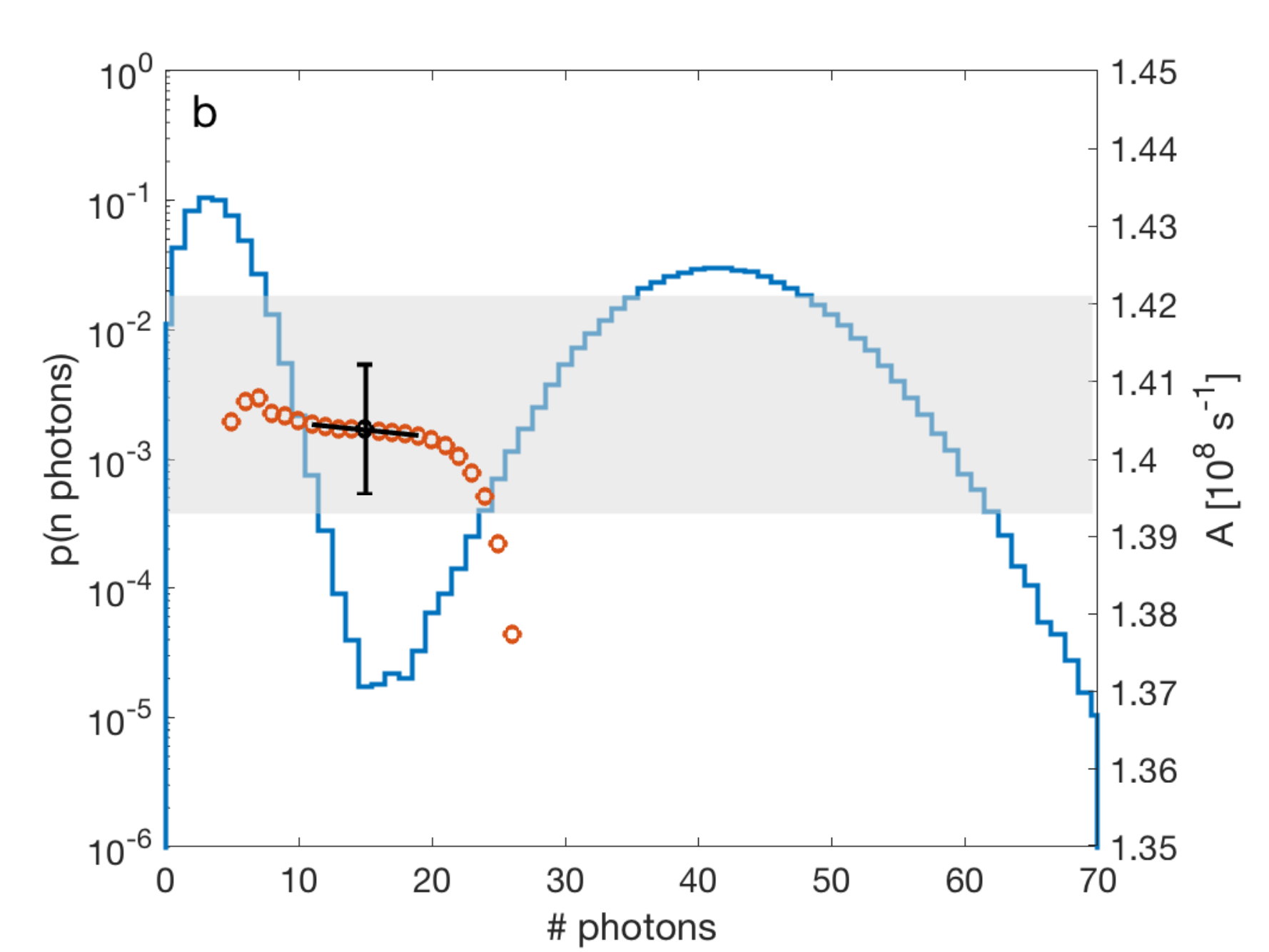}}
         \subfigure{\includegraphics[scale=0.23]{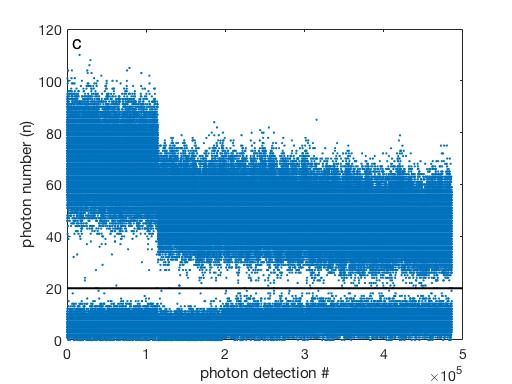}}
         \subfigure{\includegraphics[scale=0.23]{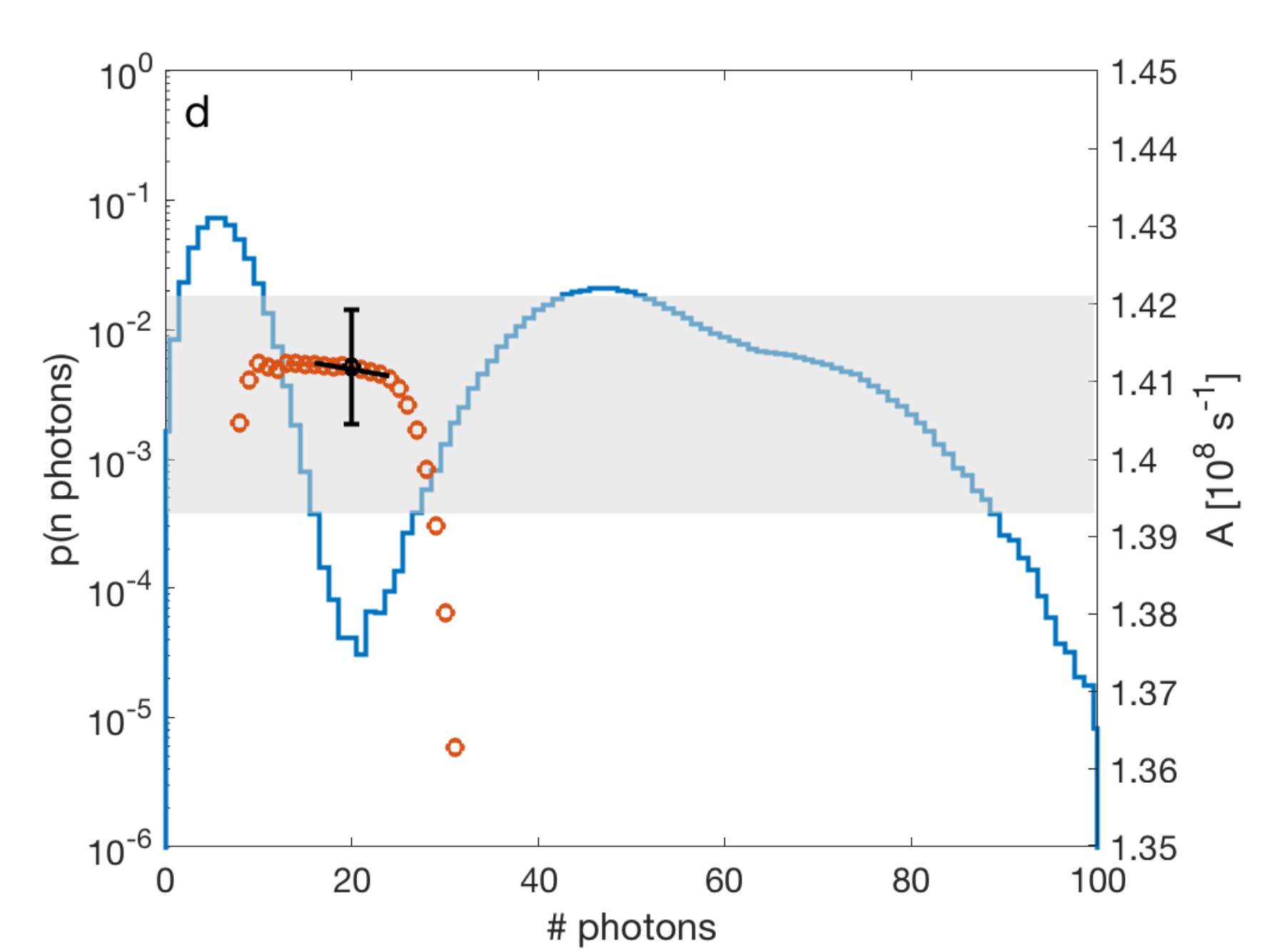}}
     \end{center}
     \caption{a) Photon number counts (blue) for an experiment (19/5, 10 hrs) with 0.5 ms detection time. The black line is the photon threshold used in the analysis. b) Histogram (blue) of the photon number count in (a). The values of the transition rates determined for different thresholds around the one used in the experiment (black symbol) are given as red symbols. A linear fit of the transition rates in an interval of $\pm3$ photons around the threshold value is shown in black. The dashed grey area represents the theoretical value and uncertainty from Ref. \cite{safronova11}. c) same as (a) for an experiment (4/6, 7 hrs) with 0.75 ms detection time. Here, the detection laser fell out of lock after ~1.75 hrs. d) Same as (b) for the photon number count in (c).}
     \label{fig:detection_err}
\end{figure}

We determined whether the ion was in a ``dark'' or ``bright'' state by counting photons ($n$) over 0.5~ms in the first experiments (17/5/19-19/5/19) and over 0.75~ms in the later ones (27/5/19-4/6/19) and setting a photon threshold ($t$) such that for $n\leq t$ the ion was considered ``dark'' while for $n>t$ the ion was considered ``bright". Photon counting traces for two experiments with two different detection times and their thresholds are shown in Figs. \ref{fig:detection_err}a,c. In the latter experiment, the cooling laser fell out of lock such that the mean ``bright'' photon number drifted during the experiment. Nevertheless, even with unlocked detection and cooling laser, no detectable systematic shifts were observed within the measurement errors.   

We determined the threshold value to minimize both ``dark'' and ``bright'' counting errors by choosing the point of lowest counting probability between the ``dark'' and ``bright'' histograms (see Fig. \ref{fig:detection_err}b,d blue trace). To quantify the effect of this threshold value on the experimental results, we calculated the dependence of the transition rate on the threshold, A($t$). The results are shown in Fig. \ref{fig:detection_err}b,d for the two different detection times. We observe that the experimental value, A($t$), is almost independent of $t$ around the chosen threshold value. There is a small linear slope of -0.0002$\cdot10^8$ s$^{-1}$/$\Delta t$ from which we estimate an uncertainty of $4\cdot10^{-4}$ due to possible error of $\pm3$ photons in the determination of the photon-count threshold. This uncertainty is small compared to our measurement uncertainty.

\subsection{Finite detection time}
In the previous section, we considered the case of the counting error due to photon statistics. This error can be reduced by increasing the detection time. However, increasing the detection time will increase the probability of a decay of the ``dark'' state into the ``bright'' state during the course of detection due to the finite lifetime of the ``dark'' state, $\tau_{\textrm{D}_{5/2}}$.

We calculated the effective decay time, $t_\textrm{eff}$, up to which a ``dark'' state is considered ``bright'', by linearizing the photon accumulation rate,
\begin{equation*}
    t_\textrm{eff}=t_\textrm{det}\frac{\bar{b}-t}{\bar{b}-\bar{d}}.
\end{equation*}
This equation accounts for the fact that with high photon threshold ``dark'' events that scatter during the detection time can still be counted correctly as ``dark'' events given that the scatter event occurred at the end of the detection period. Here, $t_\textrm{det}=0.5,0.75$ ms is the total detection time, $\bar{d}<t<\bar{b}$ is the photon threshold value introduced in the previous section and $\bar{b}$ ($\bar{d}$) are the mean ``bright'' (``dark'') photons counted in the experiment. 

The number of excessively measured ``bright'' (``dark'') photons, $\Delta b$ ($\Delta d$) is then given by,
\begin{equation*}
    \Delta b = -\Delta d = d\cdot\left(e^{t_\textrm{eff}/\tau_{\textrm{D}_{5/2}}}-1\right).    
\end{equation*}
Here, $d$ is the measured number of ``dark'' photons. For the experiment in Fig \ref{fig:detection_err}a,b we estimated 180 photons which are falsely detected as ``bright'' out of total $\sim$600,000 ``dark'' counts. For the experiment in Fig. \ref{fig:detection_err}c,d we estimated 210 false detected photons out of $\sim$470,000.

To estimate the systematic shift induced by this effect, we changed the threshold value such that 180 (210) photons were transferred from ``dark'' to ``bright'' for the two experiments. We found a systemic error of $\varepsilon=-6\cdot10^{-4}$ for both experiments. This value is small compared to our measurement uncertainty. 

\subsection{Thermal effect in AOM power stabilization}
\begin{figure}
	\centering
	\begin{center}
         \subfigure{\includegraphics[width=0.4\linewidth,trim={0cm 0cm 0cm 0cm},clip]{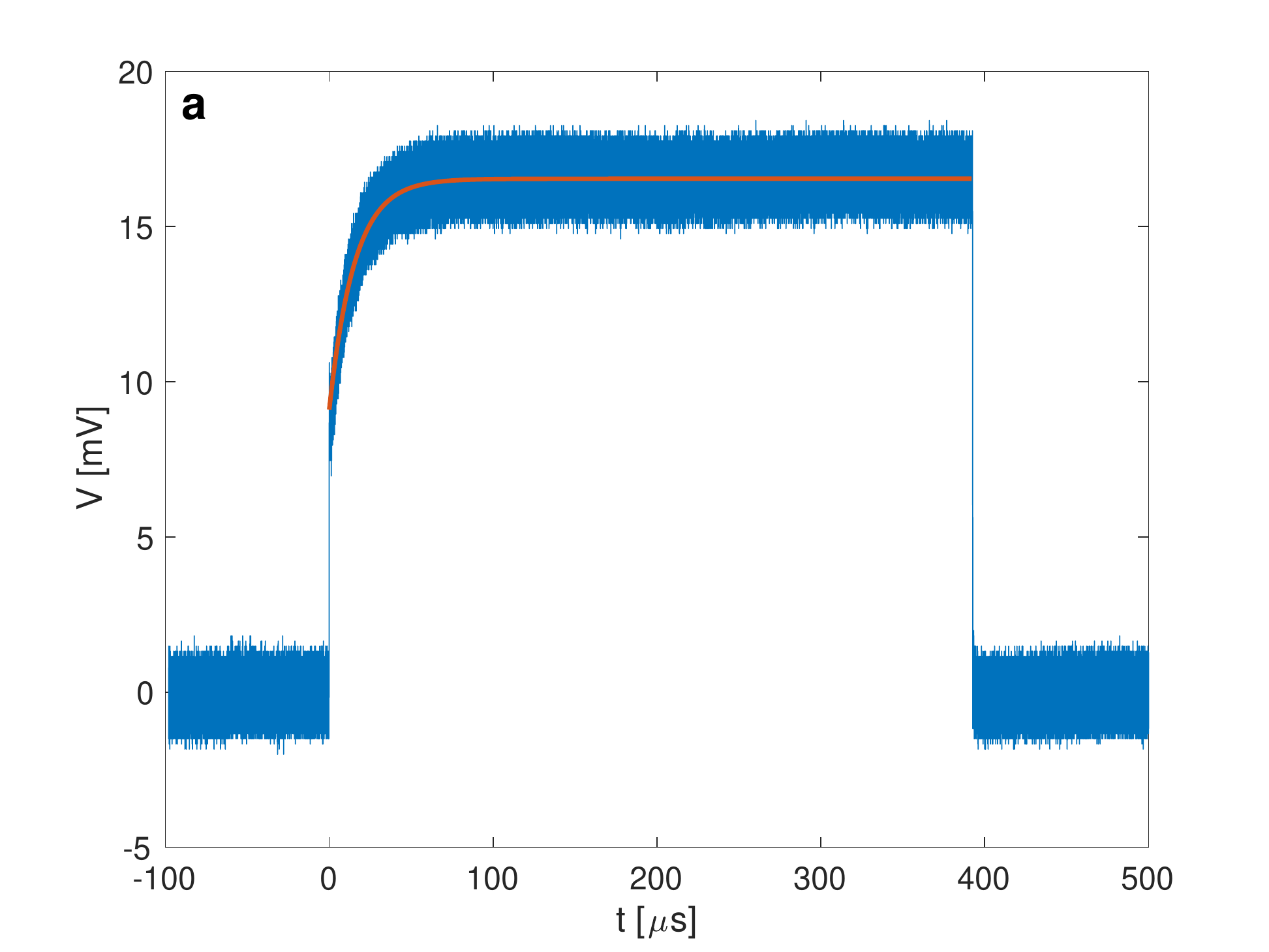}}
         \subfigure{\includegraphics[width=0.4\linewidth,trim={0cm 0cm 0cm 0cm},clip]{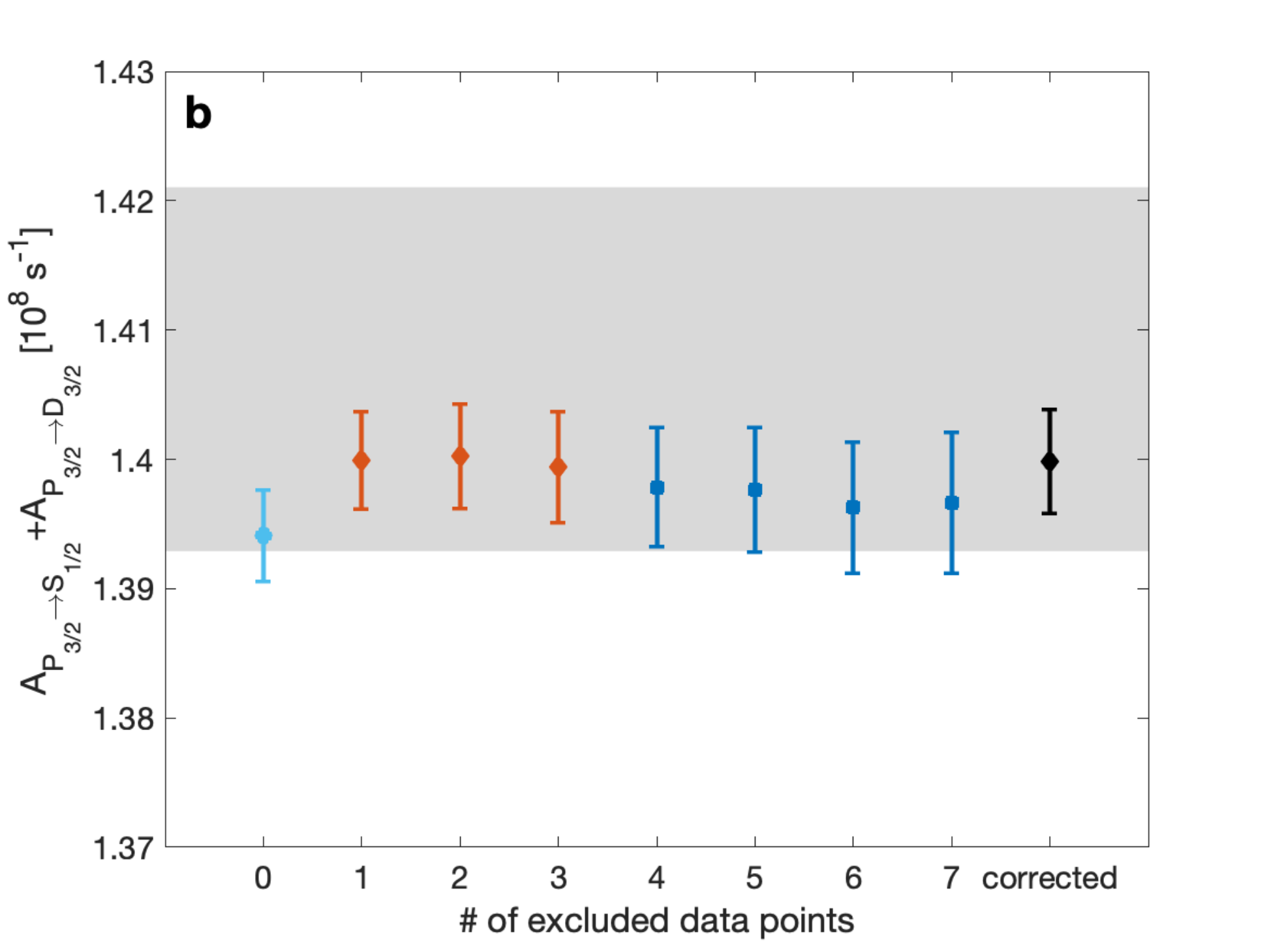}}
    \end{center}
	\caption{a) Typical snapshot of the probe-beam power measured on a fast photo-detector (blue) during a scattering-rate measurement. Red line is an exponential fit with a characteristic time of about 15 $\mu$s. b) Transition rates, $\AS+\ADt$, in which we excluded from 0 to 7 of the first data points of the scattering-rate decay curve. The results before correction reported in the main text are given in light blue circle. The values used for extracting the systematic shift are given in red diamonds where the corrected value for this systematic shift is given in black diamond. The gray-shaded area is the theoretical value of Safronova et. al. \cite{safronova11}.}
	\label{fig:lifetime_fit_bias}
\end{figure}

We used an acousto-optic modulator (AOM) to control the duration of the probe-beam pulse during the experiment. The AOM rise time is very short, typically less than a $\mu$s, however, to reach a steady-state power it takes the AOM about 15 $\mu$s (1/e). This effect is due to thermalization of the AOM crystal with the incident probe beam. In Fig. \ref{fig:lifetime_fit_bias}a we show a typical snapshot of the probe-beam power measured on a fast detector during a scattering rate measurement.

Even though the Rabi frequency cancels in the calculation of the transition rates, different effective powers between the ac-shift and the scattering rate measurements will lead to systematic errors. In the case of the ac-stark shift measurements, we applied a 2 ms D-state repump pulse using the probe-beam just before the ac-stark shift measurement began. This pulse eliminated the AOM thermal effect. We experimentally verified that there are no systematic shifts in the ac-stark shift measurement by adding a 150 $\mu$s pulse prior to the ac-stark shift measurement and comparing the resulting ac-shift with an experiment with no such pulse. The relative difference between the two measurements was 0.08(18)\% which is consistent with no shift.

On the other hand, in the case of the scattering-rate measurements, there is almost a ms delay between the D-state repump pulse and the measurement pulse due to D-shelving and state-purification pulses. For that, the AOM thermal effect is present in the scattering rate measurement and it induced a systematic shift.

To test the magnitude of this systematic shift, we analyzed the scattering-rate data excluding between 0 to 7 of the first data points of the decay curve, thus, effectively starting the scattering measurement after a time period which the AOM could reach its steady-state power. On average, each point of data we excluded amounts for roughly 25 $\mu$s of pre-AOM time. 

The results of the transition rates for this analysis are given in Fig. \ref{fig:lifetime_fit_bias}b. As expected, the transition rates value increases when excluding the first point due to the increase in the effective Rabi frequency in the measurement. The value of the transition rates remains constant when excluding from one up to three of the first data points. These values are used to calculate a systematic shift of $\varepsilon$=+0.0041. The use of less data points in the fit analysis increase the statistical error of our measurement. We quantify this as an additional error of 0.0014 to the non-corrected value. 

\subsection{Motion-induced Doppler shifts}
Mechanical effects of the ion motion affect the instantaneous detuning, $\Delta_\textrm{inst.}$, of the probe-beam light through the Doppler shift,
\begin{equation*}
    \Delta_\textrm{inst.}=\Delta+k x \omega \cos \left( \omega t\right)\equiv \Delta\left(1+\beta\cos \left( \omega t\right)\right).
\end{equation*}
Here, $k=2\pi/\lambda$ is the projection of the $k$-vector of the probe beam onto the direction of motion, $x$, with oscillation frequency $\omega$. The motion of the ion is composed of both thermal motion and micromotion with two different frequencies (700 kHz and 16.8 MHz respectively). The modulation index, $\beta=k x \omega / \Delta$, quantifies the modulation amplitude.

The duration of the scattering-rate and ac-Stark-shift measurements is much longer than one cycle of modulation. Hence, we can consider the average scattering-rate and ac-Stark-shift values,
\begin{equation*}
    \Gamma=\langle \Gamma_\textrm{inst.} \rangle = \Gamma_0 \left \langle \frac{1}{(1+\beta\cos (\omega t))^2} \right \rangle = \Gamma_0 \left (1+\frac{3}{2}\beta^2\right ),
\end{equation*}
\begin{equation*}
    \Delta E=\langle \Delta E_\textrm{inst.} \rangle = \Delta E_0 \left \langle \frac{1}{1+\beta\cos (\omega t)} \right \rangle = \Delta E_0 \left (1+\frac{1}{2}\beta^2 \right ).
\end{equation*}
Here, $\Delta E_0$ and $\Gamma_0$ are the values of the ac-Stark shift and scattering rate without the mechanical effect as given in Eqs. [1,2] of the main text. The effect of mechanical motion on the transition rates is given by,
\begin{equation*}
    \frac{\textrm{A}}{\textrm{A}_0}=\frac{1+\frac{3}{2}\beta^2}{1+\frac{1}{2}\beta^2}\approx 1+\beta^2,
\end{equation*}
thus the systematic shift due to mechanical motion is $\varepsilon=-\beta^2$. 

For the case of thermal motion, the ion is Doppler cooled to $\sim$0.5 mK such that amplitude of the thermal motion is less than 100 nm. The resulting modulation index is $\beta < 5\cdot10^{-4}$ and the systematic shift is $|\varepsilon| < 3 \cdot 10^{-7}$ which is negligible compared to our measurement uncertainty.

For the case of excess micromotion, its amplitude was compensated below our detection limit. For that, it is safe to estimate the micromotion amplitude to be smaller than 10 nm. In this case, the modulation index is $\beta < 1\cdot10^{-3}$ and the systematic shift is $|\varepsilon| < 1\cdot10^{-6}$ which is also negligible compared to our measurement uncertainty.

\subsection{Inelastic Raman scattering}
\begin{figure}
	\begin{center}
         \subfigure{\includegraphics[width=0.50\linewidth,trim={2.5cm 8cm 5cm 5cm},clip]{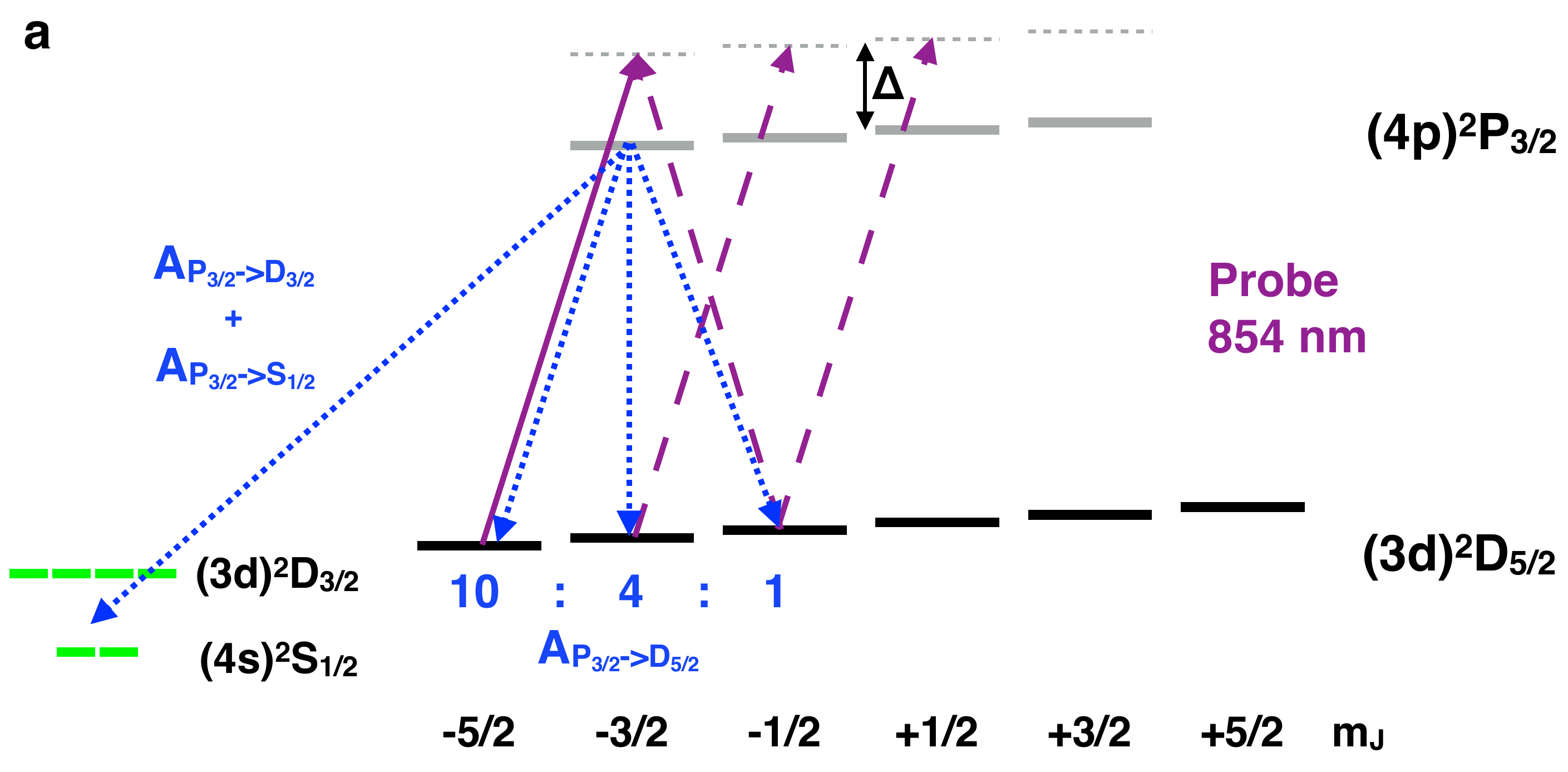}}
         \subfigure{\includegraphics[width=0.40\linewidth,trim={0cm 8cm 0cm 8cm},clip]{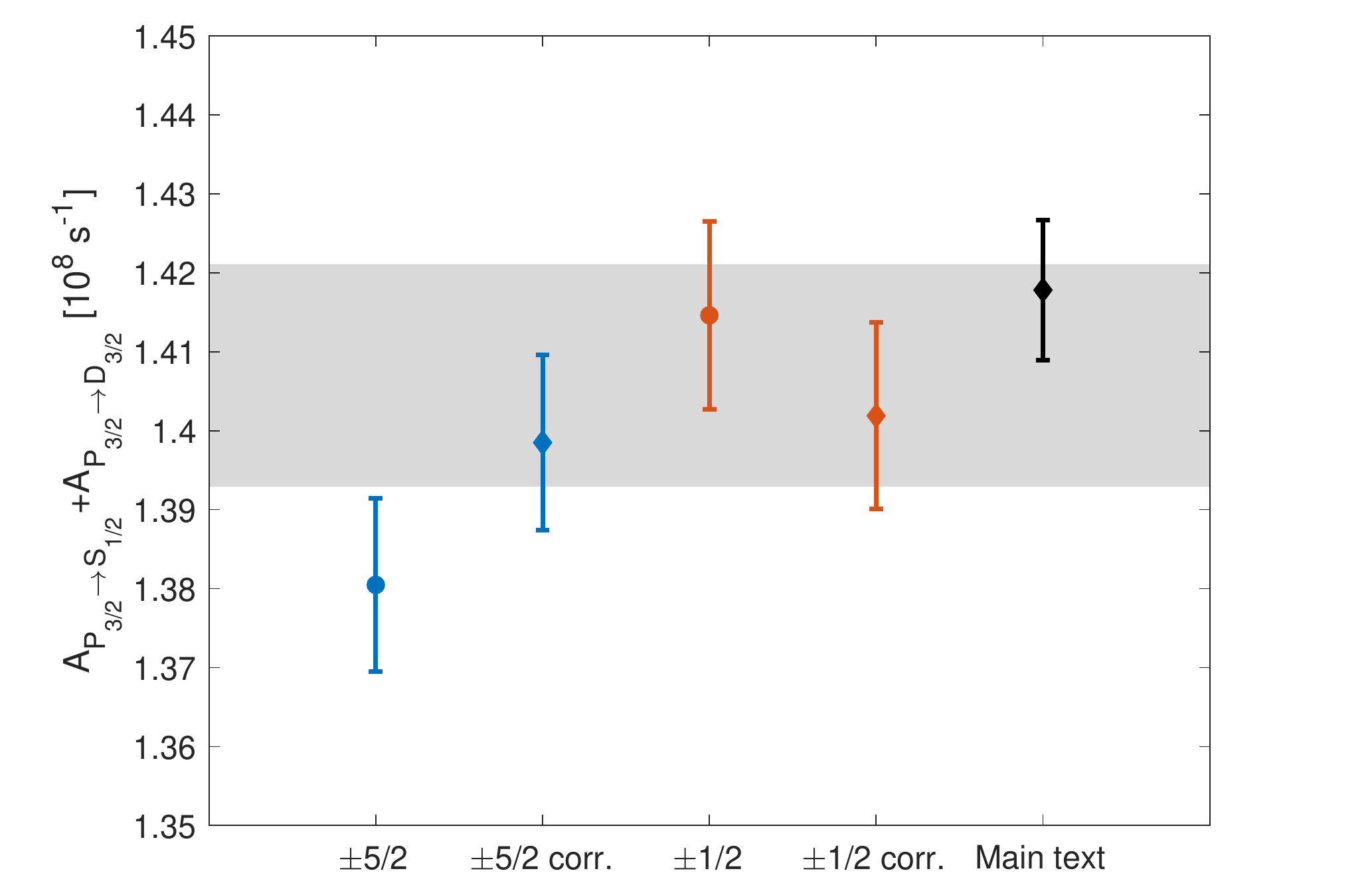}}
    \end{center}
	\caption{a) Schematic of inelastic Raman scattering from the initial D$_{5/2}$($m_D$=-5/2) state. A probe-beam (solid purple arrow) couples the D$_{5/2}$($m_D$=-5/2) state to the P$_{3/2}$($m_P$=-3/2) excited state. From this excited state, the ion can decay either to the S$_{1/2}$ or D$_{3/2}$ ``bright'' states or back to the D$_{5/2}$ ``dark'' state. The latter breaks into elastic scattering to the $m_D$=-5/2 state or inelastic scattering to the $m_D$=-3/2,-1/2 states (dotted blue arrows). The branching ratios for these events are 10:4:1 respectively. After an inelastic scattering event occurs, the probe beam couples the ion to different excited Zeeman states of the P$_{3/2}$ level (dashed purple arrows). b) Experimental verification of the inelastic process. Transition rates for an ion prepared in the D$_{5/2}$($m_D=\pm5/2$) (blue) and the D$_{5/2}$($m_D=\pm1/2$) (red) states. Circles (diamonds) represent values before (after) the correction of the systematic shift for inelastic scattering. The black diamond is the corrected value given in the main text. The gray-shaded area is the theoretical value of Safronova et. al. \cite{safronova11}.}
	\label{fig:inelastic}
\end{figure}

In the formula of the scattering rate given in Eq. [2] of the main text, we assumed that either the ion decays to ``bright'' states or it decays back to its initial Zeeman ``dark'' state (also known as elastic Rayleigh scattering). This assumption neglects the inelastic Raman scattering in which the ion can decay to different Zeeman states of the D$_{5/2}$ manifold (see Fig. \ref{fig:inelastic}a). Inelastic scattering results in the change of the Rabi frequency during the measurement instance due to different angular factors in the transition moment.

As an example (see Fig. \ref{fig:inelastic}a), we consider the case of an ion prepared in the D$_{5/2}(m=-5/2)$ state. A probe-beam with linear horizontal polarization couples this state to the P$_{3/2}(m=-3/2)$ state. From this excited state, there is a probability, $p_b=(\AS+\ADt)/(\AS+\ADt+\ADf)=0.941$, to decay to the ``bright'' states and $p_d=(1-p_b)=0.059$ probability to decay back to the D$_{5/2}$ state. In the case of decaying back to the D$_{5/2}$ state, the probability to decay to the different Zeeman states is given by,
\begin{equation*}
    p\left({\textrm{P}_{3/2}(m_P)\rightarrow\textrm{D}_{5/2}(m_D)}\right)=(2\cdot3/2+1)\cdot \tj{3/2}{1}{5/2}{-m_P}{m_P-m_D}{m_D}^2.
\end{equation*}
Here, the big brackets stand for a Wigner 3j symbol. In our example, there is a $2/3$ chance to decay back to the initial $m=-5/2$ and $4/15$ ($1/15$) chance to decay to the $m=-3/2$ ($m=-1/2$) state. The total probability for the inelastic scattering event is then given by, $p_d\cdot1/3=0.02$. Note that for the electronic ground state, S$_{1/2}$, Raman scattering is known to vanish due to destructive interference from the P$_{1/2}$ and P$_{3/2}$ states \cite{ozeri05}. In our case, due to the single transition involved, there is no such destructive interference. When the ion decays to a different Zeeman state, the Rabi frequency changes accordingly and thus the rate in which a second scattering event occurs. The Rabi frequencies for the different Zeeman states of the D$_{5/2}$ level are proportional to,
\begin{equation*}
    \Omega^2(m_D)\propto\tj{3/2}{1}{5/2}{-1-m_D}{1}{m_D}^2+\tj{3/2}{1}{5/2}{1-m_D}{-1}{m_D}^2.
\end{equation*}
In our example, $\Omega^2(-3/2)/\Omega^2(-5/2)=3/5$ and $\Omega^2(-1/2)/\Omega^2(-5/2)=2/5$.

Including the inelastic process, the decay of the ``dark'' state changes from a single exponential decay to the following expression,
\begin{equation*}
    p(\textrm{dark})=p_b e^{-\Gamma(m_D)t}+p_d\sum_{m_D'}p(\textrm{P}_{3/2}(m_P)\rightarrow\textrm{D}_{5/2}(m_D')) e^{-\Gamma(m_D')t},
\end{equation*}
with $\Gamma(m_D')=\Gamma(m_D)\cdot \Omega^2(m_D') / \Omega^2(m_D)$. For the initial Zeeman state, $m_D=\pm1/2$, the expression becomes more complicated since the probe beam initially populates two Zeeman states in the excited P$_{3/2}$ level ($m_P=\pm3/2$,$\mp1/2$). The probability to populate each of these states is given by,
\begin{equation*}
    p(m_P^\pm)=\frac{\tj{3/2}{1}{5/2}{\mp1-m_D}{\pm1}{m_D}^2}{\tj{3/2}{1}{5/2}{-1-m_D}{1}{m_D}^2+\tj{3/2}{1}{5/2}{1-m_D}{-1}{m_D}^2},
\end{equation*}
and the expression for the decay of the ``dark'' state changes accordingly,
\begin{equation*}
    p(\textrm{dark})=p_b e^{-\Gamma(m_D)t}+p_d\sum_{m_P^\pm}\sum_{m_D'}p(m_P^\pm)p(\textrm{P}_{3/2}(m_P^\pm)\rightarrow\textrm{D}_{5/2}(m_D')) e^{-\Gamma(m_D')t}.
\end{equation*}
Here, $m_P^\pm$ stands for exciting a state with Zeeman quantum number $m_P=m_D\pm1$.

The dark state population can be written in general form,
\begin{equation*}
    \textrm{p}(\textrm{``dark''})=p_\frac{5}{2}e^{-\Gamma(\frac{5}{2})(t-t_0)}+p_\frac{3}{2}e^{-\Gamma(\frac{3}{2})(t-t_0)}+p_\frac{1}{2}e^{-\Gamma(\frac{1}{2})(t-t_0)}.
\end{equation*}
Here, we used the symmetry of the Zeeman states, $\Gamma(m_D)=\Gamma(-m_D)$, and introduced back, $t_0$, which accounts for the fact that the AOM was turned on before the experiment began. The probabilities, $p_{|m|}$, indicate from which Zeeman state the ion scatters from ``dark'' to ``bright''. 
In the previous paragraph, we showed how to derive the probabilities within the approximation of a single inelastic Raman scattering event. To check our calculations and to derive more accurate probabilities, we solved the dynamical optical Bloch equations (DOBE) of our system \cite{meir17b,meir18a}. The treatment considers the 12 Zeeman levels of the S$_{1/2}$,P$_{3/2}$ and D$_{5/2}$ states (for simplicity we omitted the D$_{3/2}$ levels), a probe-beam that couples the D$_{5/2}$ and the P$_{3/2}$ states with horizontal linear polarization and all spontaneous decay channels. We initialized the density matrix in a single Zeeman state of the D$_{5/2}$ manifold and numerically calculated the density matrix evolution in time during the decay to the S$_{1/2}$ levels. We then fitted the probabilities to the DOBE numerical solution. The results of the probabilities for different initial Zeeman states using the DOBE and the single Raman scattering analytic derivation are given in Table \ref{tab:prob}.

\begin{table}
  \begin{center}
    \begin{tabular}{|l|c c c|l|}
      \hline
      %\hline
      Initial state & $p_{|\frac{5}{2}|}$ & $p_{|\frac{3}{2}|}$ &  $p_{|\frac{1}{2}|}$ & Method\\
      \hline
      \hline
      $m=\pm$5/2 & 1      & 0      & 0      & No correction \\
      \hline
      $m=\pm$5/2 & 0.9896 & 0      & 0.0104 & DOBE \\
      \hline
      $m=\pm$5/2 & 0.9802 & 0.0158 & 0.0040 & Single scattering \\
      \hline
      \hline
      $m=\pm$3/2 & 0      & 1      & 0      & No correction \\
      \hline
      $m=\pm$3/2 & 0.0160 & 0.9407 & 0.0430 & DOBE \\
      \hline
      $m=\pm$3/2 & 0      & 0.9644 & 0.0356 & Single scattering \\
      \hline
      \hline
      $m=\pm$1/2 & 0      & 0      & 1      & No correction \\
      \hline
      $m=\pm$1/2 & 0      & 0.0278 & 0.9722 & DOBE \\
      \hline
      $m=\pm$1/2 & 0.0099 & 0.0218 & 0.9684 & Single scattering \\
      \hline
      %\hline
    \end{tabular}
    \caption{Probabilities, $p_{|m|}$, of the Zeeman state before scattering from ``dark'' to ``bright'' derived from DOBE and single-scattering analytic calculations for all initial Zeeman states of the D$_{5/2}$ manifold. ``No correction" stands for the limiting case of no inelastic Raman scattering shift described by Eq. [2] of the main text.}
    \label{tab:prob}
  \end{center}
\end{table}

For the $m_D=\pm5/2,\pm3/2$ states, the inelastic Raman scattering tends to decrease the scattering rate due to pumping to states with lower Rabi frequency. 
%For that, even with our small measurement uncertainty, there was no significant difference between the $\pm5/2$ and $\pm3/2$ Zeeman states. 
The $m_D=\pm1/2$ states, however, show an increase in the scattering rate since they posses the lowest Rabi frequency. To verify the inelastic Raman effect experimentally, we performed an experiment in which the ion is prepared in the $m_D=\pm5/2$ and $m_D=\pm1/2$ states. In that experiment, we used a spectroscopy laser with different orientation with respect to the trap axis and different polarization with respect to the magnetic field axis than the one used in the original experiments to allow preparation of the ion in both the $\pm1/2,\pm5/2$ Zeeman states. The results are shown in Fig. \ref{fig:inelastic}b and are in agreement with our calculations.

\subsection{Off-resonant Raman coupling}
Since we used a linear-horizontal polarized laser beam in the experiment, we allowed for off-resonant Raman coupling between Zeeman states in the D$_{5/2}$ level which satisfy $\Delta m_D=\pm2$. This off-resonant coherent coupling dresses our initial Zeeman state with Zeeman states of $\Delta m_D=\pm2$ and thus changes the coupling to the excited P$_{3/2}$ level. 

We estimate the mixing by considering the bare-Raman coupling,
\begin{equation*}
    \Omega_\textrm{Raman}=\frac{\Omega(m_D)\Omega(m_D\pm2)}{\Delta}.
\end{equation*}
Here, $\Omega(m_D)$ is the Rabi frequency of the probe beam that couples the D$_{5/2}(m_D)$ state with the excited P$_{3/2}$ state and $\Delta$ is the detuning of the probe beam from the excited state. We estimate an upper bound for the mixing due to this coupling by considering an off-resonant Rabi flop. The average population in the coupled Zeeman state is then given by,
\begin{equation*}
    p_\textrm{mix}=\frac{1}{2}\frac{\Omega_\textrm{Raman}^2}{\Omega_\textrm{Raman}^2+\Delta_\textrm{Raman}^2}.
\end{equation*}
Here, $\Delta_\textrm{Raman}$ is the detuning between the two Zeeman states due to the external magnetic field of 4.609 Gauss. We now can calculate the upper bound for this systematic shift,
\begin{equation*}
    \varepsilon=p_\textrm{mix}\left(\frac{\Omega^2(m_D+2)}{\Omega^2(m_D)}-1\right)\leq-1.4\cdot10^{-5}.
\end{equation*}
We consider this calculated shift as an upper bound since we only included the effect on the scattering rate measurements. Similar considerations can be made for the ac-Stark shift measurements which will result in reduction of this systematic effect. 

\section{Extraction of polarizabilities}

The matrix elements that we obtained in this work as well as extracted from other measurements~\cite{gerritsma08,ramm13,hettrich15} can also be used to improve knowledge of the $4s$ and $3d_j$ polarizabilities. These quantities are of particular interest due to their relevance in the determinations of  the blackbody radiation shift in the Ca$^+$ clock \cite{safronova11,huang19}. The valence static scalar polarizability $\alpha_{0}(v)$  of an atom with one valence electron $v$ is given by
\begin{equation}
    \alpha _{\text{0}}(v)\ =\frac{2}{3(2j_{v}+1)}\sum_{k}\frac{|\langle v||D||k\rangle |^{2}}{E_{k}-E_{v}}, \label{one}
\end{equation}
where  $|\langle v||D||k\rangle$ is a reduced electric-dipole matrix element and the indices $k$ range over the $np$ states for the $4s$ electron and over the $np$ and $nf$ states for the $3d$ electron. The $4s-4p_{j}$ contributions dominate the $4s$ value so increased precision of the matrix elements improves the $4s$ polarizability. Results  obtained with the matrix elements from this work, Ref.~\cite{hettrich15} and combination of the two are listed in Columns A, B and C of Table~\ref{table5}, respectively. Theory values from \cite{safronova11} are listed for comparison. Relative uncertainties in the polarizability contributions are twice the uncertainties of the corresponding matrix elements. When values are correlated such as in the uncertainties in the $4s-4p_{1/2}$ and $4s-4p_{3/2}$ matrix elements extracted from the same work, we linearly add the uncertainties.

The differential scalar polarizability for the $4s-3d_{5/2}$ clock transition was measured in \cite{huang19} to be $-44.079(13)$~a.u. We use this value and the ground state polarizabilities from Table~\ref{table5} to extract a value of the $3d_{5/2}$ scalar polarizability, listed in the columns A, B and C last row of Table~\ref{table5}. All values are in agreement with the theory \cite{safronova11}, validating theory calculations obtained using the same method for similar systems.

\begin{table*} [th]
    \caption{\label{table5} Ca$^+$ static polarizabilities (in a.u.) obtained with the matrix elements from this work (Column A), Ref.~\cite{hettrich15} (Column B) and combination of the two (Column C). The scalar $3d_{5/2}$ polarizability in columns A,B and C is extracted by combining the resulting ground state polarizability and the differential Ca$^+$ clock polarizability $\delta\alpha_0(3d_{5/2}-4s)=-44.079(13)$~a.u. measured in \cite{huang19}. Theory values from \cite{safronova11} are listed for comparison.}
    \begin{ruledtabular}
        \begin{tabular}{lcccc}
            \multicolumn{1}{c}{}& \multicolumn{1}{c}{Theory~\cite{safronova11}}&
            \multicolumn{1}{c}{A}& \multicolumn{1}{c}{B}& \multicolumn{1}{c}{C}     \\
            \hline $4p_{1/2} - 4s$     & 24.4(2) & 24.58(15) & 24.30(7)  & 24.30(7)  \\
            $4p_{3/2} - 4s$            & 48.4(4) & 48.74(31) & 48.20(14) & 48.74(31) \\
            Other  \cite{safronova11}  & 3.36(5) & 3.36(5)   & 3.36(5)   & 3.36(5)   \\
            Total  $\alpha(4s)$        & 76.1(5) & 76.68(46) & 75.86(21) & 76.40(32) \\
            Total $\alpha_0(3d_{5/2})$ & 31.8(3) & 32.60(46) & 31.78(21) & 32.32(32) \\
        \end{tabular}
    \end{ruledtabular}
\end{table*}

We also used the $4p-3d$ matrix elements extracted in this work to evaluate $3d_j$ scalar and tensor polarizabilites, as well  as provide a consistency check of the $3d_{5/2}$ static value obtaind from the \cite{huang19} measurement that was presented in Table \ref{table5}. Tensor polarizabilities are given by
\begin{align}
    \alpha _{2}(v) & \ =(-1)^{j_{v}}\sqrt{\frac{40j_{v}(2j_{v}-1)}{3(j_{v}+1)(2j_{v}+1)(2j_{v}+3)}}\ \nonumber\\& \times \sum_{k}(-1)^{j}\left\{
    \begin{array}{lll}
        j_{v} & 1 & j \\
        1 & j_{v} & 2
    \end{array}
    \right\} \frac{|\langle v||D||k\rangle |^{2}}{E_{k}-E_{v}}, \label{two}
\end{align}
where the curly bracket stands for the Wigner 6j symbol. The results are given in Table~\ref{table6}. The scalar $3d_{5/2}$ value obtained using this method is in agreement with the results given in Table~\ref{table5}.

\begin{table*} [th]
    \caption{\label{table6} Ca$^+$ $3d$ static scalar ($\alpha_0$) and tensor ($\alpha_2$) polarizabilities (in a.u.) obtained with the matrix elements from this work (Column A), Ref.~\cite{hettrich15} (Column B) and combination of the two (Column C). The other contributions are taken from \cite{safronova11}.}
    \begin{ruledtabular}
        \begin{tabular}{llccccccccc}
            \multicolumn{1}{c}{State}&\multicolumn{1}{c}{Contr.}& \multicolumn{3}{c}{$\alpha_0$}&\multicolumn{3}{c}{$\alpha_2$}\\
            \multicolumn{2}{c}{}&
            \multicolumn{1}{c}{A}&\multicolumn{1}{c}{B}&\multicolumn{1}{c}{C} &\multicolumn{1}{c}{A}&\multicolumn{1}{c}{B}&\multicolumn{1}{c}{C} \\
            \hline
            $3d_{3/2}$& $3d_{3/2}- 4p_{1/2}$& 19.16(14)&18.97(7)&18.97(7)
                                            &-19.16(14)&-18.97(7)&-18.97(7) \\
                      & $3d_{3/2}- 4p_{3/2}$& 3.74(3)&3.71(1)&3.74(3)
                                            &2.99(3)&2.97(1)&2.99(3) \\
                      & Other  \cite{safronova11} &9.01(21)&9.01(21)&9.01(21)                                           &-1.20(4)&-1.20(4)&-1.20(4) \\
                      & Total               &31.91(27)&31.69(23)&31.72(22)
                                            &-17.37(16)&-17.20(9)&-17.18(8) \\
            \hline
            $3d_{5/2}$& $3d_{5/2}- 4p_{3/2}$& 22.69(17)&22.46(8)&
                                            &-22.69(17)&-22.46(8)& \\
                      & Other  \cite{safronova11} &9.02(17)&9.02(17)&
                                                  &-1.73(4)&-1.73(4)& \\
                   &Total                   &31.71(24)&31.48(19)&
                                            &-24.42(17)&-24.19(9)& \\
       \end{tabular}
    \end{ruledtabular} 
\end{table*}


\begin{thebibliography}{42}%
\makeatletter
\providecommand \@ifxundefined [1]{%
 \@ifx{#1\undefined}
}%
\providecommand \@ifnum [1]{%
 \ifnum #1\expandafter \@firstoftwo
 \else \expandafter \@secondoftwo
 \fi
}%
\providecommand \@ifx [1]{%
 \ifx #1\expandafter \@firstoftwo
 \else \expandafter \@secondoftwo
 \fi
}%
\providecommand \natexlab [1]{#1}%
\providecommand \enquote  [1]{``#1''}%
\providecommand \bibnamefont  [1]{#1}%
\providecommand \bibfnamefont [1]{#1}%
\providecommand \citenamefont [1]{#1}%
\providecommand \href@noop [0]{\@secondoftwo}%
\providecommand \href [0]{\begingroup \@sanitize@url \@href}%
\providecommand \@href[1]{\@@startlink{#1}\@@href}%
\providecommand \@@href[1]{\endgroup#1\@@endlink}%
\providecommand \@sanitize@url [0]{\catcode `\\12\catcode `\$12\catcode
  `\&12\catcode `\#12\catcode `\^12\catcode `\_12\catcode `\%12\relax}%
\providecommand \@@startlink[1]{}%
\providecommand \@@endlink[0]{}%
\providecommand \url  [0]{\begingroup\@sanitize@url \@url }%
\providecommand \@url [1]{\endgroup\@href {#1}{\urlprefix }}%
\providecommand \urlprefix  [0]{URL }%
\providecommand \Eprint [0]{\href }%
\providecommand \doibase [0]{http://dx.doi.org/}%
\providecommand \selectlanguage [0]{\@gobble}%
\providecommand \bibinfo  [0]{\@secondoftwo}%
\providecommand \bibfield  [0]{\@secondoftwo}%
\providecommand \translation [1]{[#1]}%
\providecommand \BibitemOpen [0]{}%
\providecommand \bibitemStop [0]{}%
\providecommand \bibitemNoStop [0]{.\EOS\space}%
\providecommand \EOS [0]{\spacefactor3000\relax}%
\providecommand \BibitemShut  [1]{\csname bibitem#1\endcsname}%
\let\auto@bib@innerbib\@empty
%</preamble>
\bibitem [{\citenamefont {Fortson}(1993)}]{fortson93}%
  \BibitemOpen
  \bibfield  {author} {\bibinfo {author} {\bibfnamefont {N.}~\bibnamefont
  {Fortson}},\ }\href@noop {} {\bibfield  {journal} {\bibinfo  {journal} {Phys.
  Rev. Lett.}\ }\textbf {\bibinfo {volume} {70}},\ \bibinfo {pages} {2383}
  (\bibinfo {year} {1993})}\BibitemShut {NoStop}%
\bibitem [{\citenamefont {Roberts}\ \emph {et~al.}(2015)\citenamefont
  {Roberts}, \citenamefont {Dzuba},\ and\ \citenamefont
  {Flambaum}}]{roberts15}%
  \BibitemOpen
  \bibfield  {author} {\bibinfo {author} {\bibfnamefont {B.}~\bibnamefont
  {Roberts}}, \bibinfo {author} {\bibfnamefont {V.}~\bibnamefont {Dzuba}}, \
  and\ \bibinfo {author} {\bibfnamefont {V.}~\bibnamefont {Flambaum}},\
  }\href@noop {} {\bibfield  {journal} {\bibinfo  {journal} {Annual Review of
  Nuclear and Particle Science}\ }\textbf {\bibinfo {volume} {65}},\ \bibinfo
  {pages} {63} (\bibinfo {year} {2015})}\BibitemShut {NoStop}%
\bibitem [{\citenamefont {Dutta}\ \emph {et~al.}(2016)\citenamefont {Dutta},
  \citenamefont {De~Munshi}, \citenamefont {Yum}, \citenamefont {Rebhi},\ and\
  \citenamefont {Mukherjee}}]{dutta16}%
  \BibitemOpen
  \bibfield  {author} {\bibinfo {author} {\bibfnamefont {T.}~\bibnamefont
  {Dutta}}, \bibinfo {author} {\bibfnamefont {D.}~\bibnamefont {De~Munshi}},
  \bibinfo {author} {\bibfnamefont {D.}~\bibnamefont {Yum}}, \bibinfo {author}
  {\bibfnamefont {R.}~\bibnamefont {Rebhi}}, \ and\ \bibinfo {author}
  {\bibfnamefont {M.}~\bibnamefont {Mukherjee}},\ }\href@noop {} {\bibfield
  {journal} {\bibinfo  {journal} {Scientific reports}\ }\textbf {\bibinfo
  {volume} {6}},\ \bibinfo {pages} {29772} (\bibinfo {year}
  {2016})}\BibitemShut {NoStop}%
\bibitem [{\citenamefont {Safronova}\ and\ \citenamefont
  {Safronova}(2011)}]{safronova11}%
  \BibitemOpen
  \bibfield  {author} {\bibinfo {author} {\bibfnamefont {M.~S.}\ \bibnamefont
  {Safronova}}\ and\ \bibinfo {author} {\bibfnamefont {U.~I.}\ \bibnamefont
  {Safronova}},\ }\href@noop {} {\bibfield  {journal} {\bibinfo  {journal}
  {Phys. Rev. A}\ }\textbf {\bibinfo {volume} {83}},\ \bibinfo {pages} {012503}
  (\bibinfo {year} {2011})}\BibitemShut {NoStop}%
\bibitem [{\citenamefont {Safronova}\ \emph {et~al.}(2011)\citenamefont
  {Safronova}, \citenamefont {Kozlov},\ and\ \citenamefont
  {Clark}}]{safronova11b}%
  \BibitemOpen
  \bibfield  {author} {\bibinfo {author} {\bibfnamefont {M.~S.}\ \bibnamefont
  {Safronova}}, \bibinfo {author} {\bibfnamefont {M.~G.}\ \bibnamefont
  {Kozlov}}, \ and\ \bibinfo {author} {\bibfnamefont {C.~W.}\ \bibnamefont
  {Clark}},\ }\href@noop {} {\bibfield  {journal} {\bibinfo  {journal} {Phys.
  Rev. Lett.}\ }\textbf {\bibinfo {volume} {107}},\ \bibinfo {pages} {143006}
  (\bibinfo {year} {2011})}\BibitemShut {NoStop}%
\bibitem [{\citenamefont {Keller}\ \emph {et~al.}(2019)\citenamefont {Keller},
  \citenamefont {Burgermeister}, \citenamefont {Kalincev}, \citenamefont
  {Didier}, \citenamefont {Kulosa}, \citenamefont {Nordmann}, \citenamefont
  {Kiethe},\ and\ \citenamefont {Mehlst{\"a}ubler}}]{keller19}%
  \BibitemOpen
  \bibfield  {author} {\bibinfo {author} {\bibfnamefont {J.}~\bibnamefont
  {Keller}}, \bibinfo {author} {\bibfnamefont {T.}~\bibnamefont
  {Burgermeister}}, \bibinfo {author} {\bibfnamefont {D.}~\bibnamefont
  {Kalincev}}, \bibinfo {author} {\bibfnamefont {A.}~\bibnamefont {Didier}},
  \bibinfo {author} {\bibfnamefont {A.~P.}\ \bibnamefont {Kulosa}}, \bibinfo
  {author} {\bibfnamefont {T.}~\bibnamefont {Nordmann}}, \bibinfo {author}
  {\bibfnamefont {J.}~\bibnamefont {Kiethe}}, \ and\ \bibinfo {author}
  {\bibfnamefont {T.~E.}\ \bibnamefont {Mehlst{\"a}ubler}},\ }\href@noop {}
  {\bibfield  {journal} {\bibinfo  {journal} {Phys. Rev. A}\ }\textbf {\bibinfo
  {volume} {99}},\ \bibinfo {pages} {013405} (\bibinfo {year}
  {2019})}\BibitemShut {NoStop}%
\bibitem [{\citenamefont {Carlsson}\ and\ \citenamefont
  {Leenaarts}(2012)}]{carlsson12}%
  \BibitemOpen
  \bibfield  {author} {\bibinfo {author} {\bibfnamefont {M.}~\bibnamefont
  {Carlsson}}\ and\ \bibinfo {author} {\bibfnamefont {J.}~\bibnamefont
  {Leenaarts}},\ }\href@noop {} {\bibfield  {journal} {\bibinfo  {journal}
  {Astronomy \& Astrophysics}\ }\textbf {\bibinfo {volume} {539}},\ \bibinfo
  {pages} {A39} (\bibinfo {year} {2012})}\BibitemShut {NoStop}%
\bibitem [{\citenamefont {Rosner}\ \emph {et~al.}(1997)\citenamefont {Rosner},
  \citenamefont {Holt},\ and\ \citenamefont {Scholl}}]{rosner97}%
  \BibitemOpen
  \bibfield  {author} {\bibinfo {author} {\bibfnamefont {S.~D.}\ \bibnamefont
  {Rosner}}, \bibinfo {author} {\bibfnamefont {R.~A.}\ \bibnamefont {Holt}}, \
  and\ \bibinfo {author} {\bibfnamefont {T.~J.}\ \bibnamefont {Scholl}},\
  }\href@noop {} {\bibfield  {journal} {\bibinfo  {journal} {Phys. Rev. A}\
  }\textbf {\bibinfo {volume} {55}},\ \bibinfo {pages} {3469} (\bibinfo {year}
  {1997})}\BibitemShut {NoStop}%
\bibitem [{\citenamefont {Jin}\ and\ \citenamefont {Church}(1993)}]{jin93}%
  \BibitemOpen
  \bibfield  {author} {\bibinfo {author} {\bibfnamefont {J.}~\bibnamefont
  {Jin}}\ and\ \bibinfo {author} {\bibfnamefont {D.~A.}\ \bibnamefont
  {Church}},\ }\href@noop {} {\bibfield  {journal} {\bibinfo  {journal} {Phys.
  Rev. Lett.}\ }\textbf {\bibinfo {volume} {70}},\ \bibinfo {pages} {3213}
  (\bibinfo {year} {1993})}\BibitemShut {NoStop}%
\bibitem [{\citenamefont {Huntemann}\ \emph {et~al.}(2016)\citenamefont
  {Huntemann}, \citenamefont {Sanner}, \citenamefont {Lipphardt}, \citenamefont
  {\mbox{Chr.} Tamm}, ,\ and\ \citenamefont {Peik}}]{huntemann16a}%
  \BibitemOpen
  \bibfield  {author} {\bibinfo {author} {\bibfnamefont {N.}~\bibnamefont
  {Huntemann}}, \bibinfo {author} {\bibfnamefont {C.}~\bibnamefont {Sanner}},
  \bibinfo {author} {\bibfnamefont {B.}~\bibnamefont {Lipphardt}}, \bibinfo
  {author} {\bibnamefont {\mbox{Chr.} Tamm}}, , \ and\ \bibinfo {author}
  {\bibfnamefont {E.}~\bibnamefont {Peik}},\ }\href@noop {} {\bibfield
  {journal} {\bibinfo  {journal} {{Phys. Rev. Lett.}}\ }\textbf {\bibinfo
  {volume} {116}},\ \bibinfo {pages} {063001} (\bibinfo {year}
  {2016})}\BibitemShut {NoStop}%
\bibitem [{\citenamefont {Brewer}\ \emph {et~al.}(2019)\citenamefont {Brewer},
  \citenamefont {Chen}, \citenamefont {Hankin}, \citenamefont {Clements},
  \citenamefont {Chou}, \citenamefont {Wineland}, \citenamefont {Hume},\ and\
  \citenamefont {Leibrandt}}]{brewer19}%
  \BibitemOpen
  \bibfield  {author} {\bibinfo {author} {\bibfnamefont {S.~M.}\ \bibnamefont
  {Brewer}}, \bibinfo {author} {\bibfnamefont {J.-S.}\ \bibnamefont {Chen}},
  \bibinfo {author} {\bibfnamefont {A.~M.}\ \bibnamefont {Hankin}}, \bibinfo
  {author} {\bibfnamefont {E.~R.}\ \bibnamefont {Clements}}, \bibinfo {author}
  {\bibfnamefont {C.~W.}\ \bibnamefont {Chou}}, \bibinfo {author}
  {\bibfnamefont {D.~J.}\ \bibnamefont {Wineland}}, \bibinfo {author}
  {\bibfnamefont {D.~B.}\ \bibnamefont {Hume}}, \ and\ \bibinfo {author}
  {\bibfnamefont {D.~R.}\ \bibnamefont {Leibrandt}},\ }\href@noop {} {\bibfield
   {journal} {\bibinfo  {journal} {Phys. Rev. Lett.}\ }\textbf {\bibinfo
  {volume} {123}},\ \bibinfo {pages} {033201} (\bibinfo {year}
  {2019})}\BibitemShut {NoStop}%
\bibitem [{\citenamefont {Monz}\ \emph {et~al.}(2016)\citenamefont {Monz},
  \citenamefont {Nigg}, \citenamefont {Martinez}, \citenamefont {Brandl},
  \citenamefont {Schindler}, \citenamefont {Rines}, \citenamefont {Wang},
  \citenamefont {Chuang},\ and\ \citenamefont {Blatt}}]{monz16a}%
  \BibitemOpen
  \bibfield  {author} {\bibinfo {author} {\bibfnamefont {T.}~\bibnamefont
  {Monz}}, \bibinfo {author} {\bibfnamefont {D.}~\bibnamefont {Nigg}}, \bibinfo
  {author} {\bibfnamefont {E.~A.}\ \bibnamefont {Martinez}}, \bibinfo {author}
  {\bibfnamefont {M.~F.}\ \bibnamefont {Brandl}}, \bibinfo {author}
  {\bibfnamefont {P.}~\bibnamefont {Schindler}}, \bibinfo {author}
  {\bibfnamefont {R.}~\bibnamefont {Rines}}, \bibinfo {author} {\bibfnamefont
  {S.~X.}\ \bibnamefont {Wang}}, \bibinfo {author} {\bibfnamefont {I.~L.}\
  \bibnamefont {Chuang}}, \ and\ \bibinfo {author} {\bibfnamefont
  {R.}~\bibnamefont {Blatt}},\ }\href {\doibase 10.1126/science.aad9480}
  {\bibfield  {journal} {\bibinfo  {journal} {Science}\ }\textbf {\bibinfo
  {volume} {351}},\ \bibinfo {pages} {1068} (\bibinfo {year} {2016})},\ \Eprint
  {http://arxiv.org/abs/1507.08852} {1507.08852} \BibitemShut {NoStop}%
\bibitem [{\citenamefont {Linke}\ \emph {et~al.}(2017)\citenamefont {Linke},
  \citenamefont {Maslov}, \citenamefont {Roetteler}, \citenamefont {Debnath},
  \citenamefont {Figgatt}, \citenamefont {Landsman}, \citenamefont {Wright},\
  and\ \citenamefont {Monroe}}]{linke17a}%
  \BibitemOpen
  \bibfield  {author} {\bibinfo {author} {\bibfnamefont {N.~M.}\ \bibnamefont
  {Linke}}, \bibinfo {author} {\bibfnamefont {D.}~\bibnamefont {Maslov}},
  \bibinfo {author} {\bibfnamefont {M.}~\bibnamefont {Roetteler}}, \bibinfo
  {author} {\bibfnamefont {S.}~\bibnamefont {Debnath}}, \bibinfo {author}
  {\bibfnamefont {C.}~\bibnamefont {Figgatt}}, \bibinfo {author} {\bibfnamefont
  {K.~A.}\ \bibnamefont {Landsman}}, \bibinfo {author} {\bibfnamefont
  {K.}~\bibnamefont {Wright}}, \ and\ \bibinfo {author} {\bibfnamefont
  {C.}~\bibnamefont {Monroe}},\ }\href {\doibase 10.1073/pnas.1618020114}
  {\bibfield  {journal} {\bibinfo  {journal} {Proc. Nat. Acad. Sci. USA}\
  }\textbf {\bibinfo {volume} {114}},\ \bibinfo {pages} {3305} (\bibinfo {year}
  {2017})},\ \Eprint {http://arxiv.org/abs/1702.01852} {1702.01852}
  \BibitemShut {NoStop}%
\bibitem [{\citenamefont {Hettrich}\ \emph {et~al.}(2015)\citenamefont
  {Hettrich}, \citenamefont {Ruster}, \citenamefont {Kaufmann}, \citenamefont
  {Roos}, \citenamefont {Schmiegelow}, \citenamefont {Schmidt-Kaler},\ and\
  \citenamefont {Poschinger}}]{hettrich15}%
  \BibitemOpen
  \bibfield  {author} {\bibinfo {author} {\bibfnamefont {M.}~\bibnamefont
  {Hettrich}}, \bibinfo {author} {\bibfnamefont {T.}~\bibnamefont {Ruster}},
  \bibinfo {author} {\bibfnamefont {H.}~\bibnamefont {Kaufmann}}, \bibinfo
  {author} {\bibfnamefont {C.~F.}\ \bibnamefont {Roos}}, \bibinfo {author}
  {\bibfnamefont {C.~T.}\ \bibnamefont {Schmiegelow}}, \bibinfo {author}
  {\bibfnamefont {F.}~\bibnamefont {Schmidt-Kaler}}, \ and\ \bibinfo {author}
  {\bibfnamefont {U.~G.}\ \bibnamefont {Poschinger}},\ }\href@noop {}
  {\bibfield  {journal} {\bibinfo  {journal} {Phys. Rev. Lett.}\ }\textbf
  {\bibinfo {volume} {115}},\ \bibinfo {pages} {143003} (\bibinfo {year}
  {2015})}\BibitemShut {NoStop}%
\bibitem [{\citenamefont {Gerritsma}\ \emph {et~al.}(2008)\citenamefont
  {Gerritsma}, \citenamefont {Kirchmair}, \citenamefont {Z{\"a}hringer},
  \citenamefont {Benhelm}, \citenamefont {Blatt},\ and\ \citenamefont
  {Roos}}]{gerritsma08}%
  \BibitemOpen
  \bibfield  {author} {\bibinfo {author} {\bibfnamefont {R.}~\bibnamefont
  {Gerritsma}}, \bibinfo {author} {\bibfnamefont {G.}~\bibnamefont
  {Kirchmair}}, \bibinfo {author} {\bibfnamefont {F.}~\bibnamefont
  {Z{\"a}hringer}}, \bibinfo {author} {\bibfnamefont {J.}~\bibnamefont
  {Benhelm}}, \bibinfo {author} {\bibfnamefont {R.}~\bibnamefont {Blatt}}, \
  and\ \bibinfo {author} {\bibfnamefont {C.}~\bibnamefont {Roos}},\ }\href@noop
  {} {\bibfield  {journal} {\bibinfo  {journal} {The European Physical Journal
  D}\ }\textbf {\bibinfo {volume} {50}},\ \bibinfo {pages} {13} (\bibinfo
  {year} {2008})}\BibitemShut {NoStop}%
\bibitem [{\citenamefont {Ramm}\ \emph {et~al.}(2013)\citenamefont {Ramm},
  \citenamefont {Pruttivarasin}, \citenamefont {Kokish}, \citenamefont
  {Talukdar},\ and\ \citenamefont {H{\"a}ffner}}]{ramm13}%
  \BibitemOpen
  \bibfield  {author} {\bibinfo {author} {\bibfnamefont {M.}~\bibnamefont
  {Ramm}}, \bibinfo {author} {\bibfnamefont {T.}~\bibnamefont {Pruttivarasin}},
  \bibinfo {author} {\bibfnamefont {M.}~\bibnamefont {Kokish}}, \bibinfo
  {author} {\bibfnamefont {I.}~\bibnamefont {Talukdar}}, \ and\ \bibinfo
  {author} {\bibfnamefont {H.}~\bibnamefont {H{\"a}ffner}},\ }\href@noop {}
  {\bibfield  {journal} {\bibinfo  {journal} {Physical review letters}\
  }\textbf {\bibinfo {volume} {111}},\ \bibinfo {pages} {023004} (\bibinfo
  {year} {2013})}\BibitemShut {NoStop}%
\bibitem [{\citenamefont {Arnold}\ \emph {et~al.}(2019)\citenamefont {Arnold},
  \citenamefont {Kaewuam}, \citenamefont {Tan}, \citenamefont {Porsev},
  \citenamefont {Safronova},\ and\ \citenamefont {Barrett}}]{arnold19}%
  \BibitemOpen
  \bibfield  {author} {\bibinfo {author} {\bibfnamefont {K.~J.}\ \bibnamefont
  {Arnold}}, \bibinfo {author} {\bibfnamefont {R.}~\bibnamefont {Kaewuam}},
  \bibinfo {author} {\bibfnamefont {T.~R.}\ \bibnamefont {Tan}}, \bibinfo
  {author} {\bibfnamefont {S.~G.}\ \bibnamefont {Porsev}}, \bibinfo {author}
  {\bibfnamefont {M.~S.}\ \bibnamefont {Safronova}}, \ and\ \bibinfo {author}
  {\bibfnamefont {M.~D.}\ \bibnamefont {Barrett}},\ }\href@noop {} {\bibfield
  {journal} {\bibinfo  {journal} {Phys. Rev. A}\ }\textbf {\bibinfo {volume}
  {99}},\ \bibinfo {pages} {012510} (\bibinfo {year} {2019})}\BibitemShut
  {NoStop}%
\bibitem [{sm()}]{sm}%
  \BibitemOpen
  \href@noop {} {}\bibinfo {note} {Supplemental material is available
  online.}\BibitemShut {Stop}%
\bibitem [{\citenamefont {Meir}\ \emph {et~al.}(2019)\citenamefont {Meir},
  \citenamefont {Hegi}, \citenamefont {Najafian}, \citenamefont {Sinhal},\ and\
  \citenamefont {Willitsch}}]{meir19a}%
  \BibitemOpen
  \bibfield  {author} {\bibinfo {author} {\bibfnamefont {Z.}~\bibnamefont
  {Meir}}, \bibinfo {author} {\bibfnamefont {G.}~\bibnamefont {Hegi}}, \bibinfo
  {author} {\bibfnamefont {K.}~\bibnamefont {Najafian}}, \bibinfo {author}
  {\bibfnamefont {M.}~\bibnamefont {Sinhal}}, \ and\ \bibinfo {author}
  {\bibfnamefont {S.}~\bibnamefont {Willitsch}},\ }\href@noop {} {\bibfield
  {journal} {\bibinfo  {journal} {Faraday Discuss.}\ }\textbf {\bibinfo
  {volume} {217}},\ \bibinfo {pages} {561} (\bibinfo {year}
  {2019})}\BibitemShut {NoStop}%
\bibitem [{\citenamefont {Ozeri}\ \emph {et~al.}(2005)\citenamefont {Ozeri},
  \citenamefont {Langer}, \citenamefont {Jost}, \citenamefont {DeMarco},
  \citenamefont {Ben-Kish}, \citenamefont {Blakestad}, \citenamefont {Britton},
  \citenamefont {Chiaverini}, \citenamefont {Itano}, \citenamefont {Hume},
  \citenamefont {Leibfried}, \citenamefont {Rosenband}, \citenamefont
  {Schmidt},\ and\ \citenamefont {Wineland}}]{ozeri05}%
  \BibitemOpen
  \bibfield  {author} {\bibinfo {author} {\bibfnamefont {R.}~\bibnamefont
  {Ozeri}}, \bibinfo {author} {\bibfnamefont {C.}~\bibnamefont {Langer}},
  \bibinfo {author} {\bibfnamefont {J.~D.}\ \bibnamefont {Jost}}, \bibinfo
  {author} {\bibfnamefont {B.~L.}\ \bibnamefont {DeMarco}}, \bibinfo {author}
  {\bibfnamefont {A.}~\bibnamefont {Ben-Kish}}, \bibinfo {author}
  {\bibfnamefont {B.~R.}\ \bibnamefont {Blakestad}}, \bibinfo {author}
  {\bibfnamefont {J.}~\bibnamefont {Britton}}, \bibinfo {author} {\bibfnamefont
  {J.}~\bibnamefont {Chiaverini}}, \bibinfo {author} {\bibfnamefont {W.~M.}\
  \bibnamefont {Itano}}, \bibinfo {author} {\bibfnamefont {D.}~\bibnamefont
  {Hume}}, \bibinfo {author} {\bibfnamefont {D.}~\bibnamefont {Leibfried}},
  \bibinfo {author} {\bibfnamefont {T.}~\bibnamefont {Rosenband}}, \bibinfo
  {author} {\bibfnamefont {P.}~\bibnamefont {Schmidt}}, \ and\ \bibinfo
  {author} {\bibfnamefont {D.~J.}\ \bibnamefont {Wineland}},\ }\href@noop {}
  {\bibfield  {journal} {\bibinfo  {journal} {Phys. Rev. Lett.}\ }\textbf
  {\bibinfo {volume} {95}},\ \bibinfo {pages} {030403} (\bibinfo {year}
  {2005})}\BibitemShut {NoStop}%
\bibitem [{\citenamefont {Meir}\ \emph {et~al.}(2017)\citenamefont {Meir},
  \citenamefont {Sikorsky}, \citenamefont {Akerman}, \citenamefont
  {Ben-shlomi}, \citenamefont {Pinkas},\ and\ \citenamefont {Ozeri}}]{meir17b}%
  \BibitemOpen
  \bibfield  {author} {\bibinfo {author} {\bibfnamefont {Z.}~\bibnamefont
  {Meir}}, \bibinfo {author} {\bibfnamefont {T.}~\bibnamefont {Sikorsky}},
  \bibinfo {author} {\bibfnamefont {N.}~\bibnamefont {Akerman}}, \bibinfo
  {author} {\bibfnamefont {R.}~\bibnamefont {Ben-shlomi}}, \bibinfo {author}
  {\bibfnamefont {M.}~\bibnamefont {Pinkas}}, \ and\ \bibinfo {author}
  {\bibfnamefont {R.}~\bibnamefont {Ozeri}},\ }\href@noop {} {\bibfield
  {journal} {\bibinfo  {journal} {Phys. Rev. A}\ }\textbf {\bibinfo {volume}
  {96}},\ \bibinfo {pages} {020701(R)} (\bibinfo {year} {2017})}\BibitemShut
  {NoStop}%
\bibitem [{\citenamefont {Meir}\ \emph {et~al.}(2018)\citenamefont {Meir},
  \citenamefont {Pinkas}, \citenamefont {Sikorsky}, \citenamefont {Ben-shlomi},
  \citenamefont {Akerman},\ and\ \citenamefont {Ozeri}}]{meir18a}%
  \BibitemOpen
  \bibfield  {author} {\bibinfo {author} {\bibfnamefont {Z.}~\bibnamefont
  {Meir}}, \bibinfo {author} {\bibfnamefont {M.}~\bibnamefont {Pinkas}},
  \bibinfo {author} {\bibfnamefont {T.}~\bibnamefont {Sikorsky}}, \bibinfo
  {author} {\bibfnamefont {R.}~\bibnamefont {Ben-shlomi}}, \bibinfo {author}
  {\bibfnamefont {N.}~\bibnamefont {Akerman}}, \ and\ \bibinfo {author}
  {\bibfnamefont {R.}~\bibnamefont {Ozeri}},\ }\href@noop {} {\bibfield
  {journal} {\bibinfo  {journal} {Phys. Rev. Lett.}\ }\textbf {\bibinfo
  {volume} {121}},\ \bibinfo {pages} {053402} (\bibinfo {year}
  {2018})}\BibitemShut {NoStop}%
\bibitem [{\citenamefont {Sahoo}\ \emph {et~al.}(2009)\citenamefont {Sahoo},
  \citenamefont {Das},\ and\ \citenamefont {Mukherjee}}]{sahoo09}%
  \BibitemOpen
  \bibfield  {author} {\bibinfo {author} {\bibfnamefont {B.~K.}\ \bibnamefont
  {Sahoo}}, \bibinfo {author} {\bibfnamefont {B.~P.}\ \bibnamefont {Das}}, \
  and\ \bibinfo {author} {\bibfnamefont {D.}~\bibnamefont {Mukherjee}},\
  }\href@noop {} {\bibfield  {journal} {\bibinfo  {journal} {Phys. Rev. A}\
  }\textbf {\bibinfo {volume} {79}},\ \bibinfo {pages} {052511} (\bibinfo
  {year} {2009})}\BibitemShut {NoStop}%
\bibitem [{\citenamefont {Smith}\ and\ \citenamefont
  {Gallagher}(1966)}]{smith66}%
  \BibitemOpen
  \bibfield  {author} {\bibinfo {author} {\bibfnamefont {W.~W.}\ \bibnamefont
  {Smith}}\ and\ \bibinfo {author} {\bibfnamefont {A.}~\bibnamefont
  {Gallagher}},\ }\href@noop {} {\bibfield  {journal} {\bibinfo  {journal}
  {Phys. Rev.}\ }\textbf {\bibinfo {volume} {145}},\ \bibinfo {pages} {26}
  (\bibinfo {year} {1966})}\BibitemShut {NoStop}%
\bibitem [{\citenamefont {Rambow}\ and\ \citenamefont
  {Schearer}(1976)}]{rambow76}%
  \BibitemOpen
  \bibfield  {author} {\bibinfo {author} {\bibfnamefont {F.}~\bibnamefont
  {Rambow}}\ and\ \bibinfo {author} {\bibfnamefont {L.}~\bibnamefont
  {Schearer}},\ }\href@noop {} {\bibfield  {journal} {\bibinfo  {journal}
  {Phys. Rev. A}\ }\textbf {\bibinfo {volume} {14}},\ \bibinfo {pages} {1735}
  (\bibinfo {year} {1976})}\BibitemShut {NoStop}%
\bibitem [{\citenamefont {Ansbacher}\ \emph {et~al.}(1985)\citenamefont
  {Ansbacher}, \citenamefont {Inamdar},\ and\ \citenamefont
  {Pinnington}}]{ansbacher85}%
  \BibitemOpen
  \bibfield  {author} {\bibinfo {author} {\bibfnamefont {W.}~\bibnamefont
  {Ansbacher}}, \bibinfo {author} {\bibfnamefont {A.}~\bibnamefont {Inamdar}},
  \ and\ \bibinfo {author} {\bibfnamefont {E.}~\bibnamefont {Pinnington}},\
  }\href@noop {} {\bibfield  {journal} {\bibinfo  {journal} {Phys. Lett. A}\
  }\textbf {\bibinfo {volume} {110}},\ \bibinfo {pages} {383} (\bibinfo {year}
  {1985})}\BibitemShut {NoStop}%
\bibitem [{\citenamefont {Gosselin}\ \emph {et~al.}(1988)\citenamefont
  {Gosselin}, \citenamefont {Pinnington},\ and\ \citenamefont
  {Ansbacher}}]{gosselin88}%
  \BibitemOpen
  \bibfield  {author} {\bibinfo {author} {\bibfnamefont {R.~N.}\ \bibnamefont
  {Gosselin}}, \bibinfo {author} {\bibfnamefont {E.~H.}\ \bibnamefont
  {Pinnington}}, \ and\ \bibinfo {author} {\bibfnamefont {W.}~\bibnamefont
  {Ansbacher}},\ }\href@noop {} {\bibfield  {journal} {\bibinfo  {journal}
  {Phys. Rev. A}\ }\textbf {\bibinfo {volume} {38}},\ \bibinfo {pages} {4887}
  (\bibinfo {year} {1988})}\BibitemShut {NoStop}%
\bibitem [{\citenamefont {Theodosiou}(1989)}]{theodosiou89}%
  \BibitemOpen
  \bibfield  {author} {\bibinfo {author} {\bibfnamefont {C.~E.}\ \bibnamefont
  {Theodosiou}},\ }\href@noop {} {\bibfield  {journal} {\bibinfo  {journal}
  {Phys. Rev. A}\ }\textbf {\bibinfo {volume} {39}},\ \bibinfo {pages} {4880}
  (\bibinfo {year} {1989})}\BibitemShut {NoStop}%
\bibitem [{\citenamefont {Vaeck}\ \emph {et~al.}(1992)\citenamefont {Vaeck},
  \citenamefont {Godefroid},\ and\ \citenamefont {FroeseFischer}}]{vaeck92}%
  \BibitemOpen
  \bibfield  {author} {\bibinfo {author} {\bibfnamefont {N.}~\bibnamefont
  {Vaeck}}, \bibinfo {author} {\bibfnamefont {M.}~\bibnamefont {Godefroid}}, \
  and\ \bibinfo {author} {\bibfnamefont {C.}~\bibnamefont {FroeseFischer}},\
  }\href@noop {} {\bibfield  {journal} {\bibinfo  {journal} {Phys. Rev. A}\
  }\textbf {\bibinfo {volume} {46}},\ \bibinfo {pages} {3704} (\bibinfo {year}
  {1992})}\BibitemShut {NoStop}%
\bibitem [{\citenamefont {Liaw}(1995)}]{liaw95}%
  \BibitemOpen
  \bibfield  {author} {\bibinfo {author} {\bibfnamefont {S.-S.}\ \bibnamefont
  {Liaw}},\ }\href@noop {} {\bibfield  {journal} {\bibinfo  {journal} {Phys.
  Rev. A}\ }\textbf {\bibinfo {volume} {51}},\ \bibinfo {pages} {R1723}
  (\bibinfo {year} {1995})}\BibitemShut {NoStop}%
\bibitem [{tw()}]{tw}%
  \BibitemOpen
  \href@noop {} {}\bibinfo {note} {This work.}\BibitemShut {Stop}%
\bibitem [{\citenamefont {Huang}\ \emph {et~al.}(2019)\citenamefont {Huang},
  \citenamefont {Guan}, \citenamefont {Zeng}, \citenamefont {Tang},\ and\
  \citenamefont {Gao}}]{huang19}%
  \BibitemOpen
  \bibfield  {author} {\bibinfo {author} {\bibfnamefont {Y.}~\bibnamefont
  {Huang}}, \bibinfo {author} {\bibfnamefont {H.}~\bibnamefont {Guan}},
  \bibinfo {author} {\bibfnamefont {M.}~\bibnamefont {Zeng}}, \bibinfo {author}
  {\bibfnamefont {L.}~\bibnamefont {Tang}}, \ and\ \bibinfo {author}
  {\bibfnamefont {K.}~\bibnamefont {Gao}},\ }\href@noop {} {\bibfield
  {journal} {\bibinfo  {journal} {Physical Review A}\ }\textbf {\bibinfo
  {volume} {99}},\ \bibinfo {pages} {011401} (\bibinfo {year}
  {2019})}\BibitemShut {NoStop}%
\bibitem [{\citenamefont {Wolf}\ \emph {et~al.}(2016)\citenamefont {Wolf},
  \citenamefont {Wan}, \citenamefont {Heip}, \citenamefont {Gebert},
  \citenamefont {Shi},\ and\ \citenamefont {Schmidt}}]{wolf16a}%
  \BibitemOpen
  \bibfield  {author} {\bibinfo {author} {\bibfnamefont {F.}~\bibnamefont
  {Wolf}}, \bibinfo {author} {\bibfnamefont {Y.}~\bibnamefont {Wan}}, \bibinfo
  {author} {\bibfnamefont {J.~C.}\ \bibnamefont {Heip}}, \bibinfo {author}
  {\bibfnamefont {F.}~\bibnamefont {Gebert}}, \bibinfo {author} {\bibfnamefont
  {C.}~\bibnamefont {Shi}}, \ and\ \bibinfo {author} {\bibfnamefont {P.~O.}\
  \bibnamefont {Schmidt}},\ }\href@noop {} {\bibfield  {journal} {\bibinfo
  {journal} {Nature}\ }\textbf {\bibinfo {volume} {530}},\ \bibinfo {pages}
  {457} (\bibinfo {year} {2016})}\BibitemShut {NoStop}%
\bibitem [{\citenamefont {Chou}\ \emph {et~al.}(2017)\citenamefont {Chou},
  \citenamefont {Kurz}, \citenamefont {Hume}, \citenamefont {Plessow},
  \citenamefont {Leibrandt},\ and\ \citenamefont {Leibfried}}]{chou17a}%
  \BibitemOpen
  \bibfield  {author} {\bibinfo {author} {\bibfnamefont {C.-w.}\ \bibnamefont
  {Chou}}, \bibinfo {author} {\bibfnamefont {C.}~\bibnamefont {Kurz}}, \bibinfo
  {author} {\bibfnamefont {D.~B.}\ \bibnamefont {Hume}}, \bibinfo {author}
  {\bibfnamefont {P.~N.}\ \bibnamefont {Plessow}}, \bibinfo {author}
  {\bibfnamefont {D.~R.}\ \bibnamefont {Leibrandt}}, \ and\ \bibinfo {author}
  {\bibfnamefont {D.}~\bibnamefont {Leibfried}},\ }\href@noop {} {\bibfield
  {journal} {\bibinfo  {journal} {Nature}\ }\textbf {\bibinfo {volume} {545}},\
  \bibinfo {pages} {203} (\bibinfo {year} {2017})}\BibitemShut {NoStop}%
\bibitem [{\citenamefont {Tong}\ \emph {et~al.}(2010)\citenamefont {Tong},
  \citenamefont {Winney},\ and\ \citenamefont {Willitsch}}]{tong10a}%
  \BibitemOpen
  \bibfield  {author} {\bibinfo {author} {\bibfnamefont {X.}~\bibnamefont
  {Tong}}, \bibinfo {author} {\bibfnamefont {A.~H.}\ \bibnamefont {Winney}}, \
  and\ \bibinfo {author} {\bibfnamefont {S.}~\bibnamefont {Willitsch}},\
  }\href@noop {} {\bibfield  {journal} {\bibinfo  {journal} {Phys. Rev. Lett.}\
  }\textbf {\bibinfo {volume} {105}},\ \bibinfo {pages} {143001} (\bibinfo
  {year} {2010})}\BibitemShut {NoStop}%
\bibitem [{\citenamefont {Germann}\ \emph {et~al.}(2014)\citenamefont
  {Germann}, \citenamefont {Tong},\ and\ \citenamefont
  {Willitsch}}]{germann14a}%
  \BibitemOpen
  \bibfield  {author} {\bibinfo {author} {\bibfnamefont {M.}~\bibnamefont
  {Germann}}, \bibinfo {author} {\bibfnamefont {X.}~\bibnamefont {Tong}}, \
  and\ \bibinfo {author} {\bibfnamefont {S.}~\bibnamefont {Willitsch}},\
  }\href@noop {} {\bibfield  {journal} {\bibinfo  {journal} {Nat. Phys.}\
  }\textbf {\bibinfo {volume} {10}},\ \bibinfo {pages} {820} (\bibinfo {year}
  {2014})}\BibitemShut {NoStop}%
\bibitem [{\citenamefont {Koelemeij}\ \emph {et~al.}(2007)\citenamefont
  {Koelemeij}, \citenamefont {Roth},\ and\ \citenamefont
  {Schiller}}]{koelemeij07a}%
  \BibitemOpen
  \bibfield  {author} {\bibinfo {author} {\bibfnamefont {J.~C.~J.}\
  \bibnamefont {Koelemeij}}, \bibinfo {author} {\bibfnamefont {B.}~\bibnamefont
  {Roth}}, \ and\ \bibinfo {author} {\bibfnamefont {S.}~\bibnamefont
  {Schiller}},\ }\href@noop {} {\bibfield  {journal} {\bibinfo  {journal}
  {Phys. Rev.~A}\ }\textbf {\bibinfo {volume} {76}},\ \bibinfo {pages} {023413}
  (\bibinfo {year} {2007})}\BibitemShut {NoStop}%
\bibitem [{\citenamefont {Hume}\ \emph {et~al.}(2011)\citenamefont {Hume},
  \citenamefont {Chou}, \citenamefont {Leibrandt}, \citenamefont {Thorpe},
  \citenamefont {Wineland},\ and\ \citenamefont {Rosenband}}]{hume11a}%
  \BibitemOpen
  \bibfield  {author} {\bibinfo {author} {\bibfnamefont {D.~B.}\ \bibnamefont
  {Hume}}, \bibinfo {author} {\bibfnamefont {C.~W.}\ \bibnamefont {Chou}},
  \bibinfo {author} {\bibfnamefont {D.~R.}\ \bibnamefont {Leibrandt}}, \bibinfo
  {author} {\bibfnamefont {M.~J.}\ \bibnamefont {Thorpe}}, \bibinfo {author}
  {\bibfnamefont {D.~J.}\ \bibnamefont {Wineland}}, \ and\ \bibinfo {author}
  {\bibfnamefont {T.}~\bibnamefont {Rosenband}},\ }\href@noop {} {\bibfield
  {journal} {\bibinfo  {journal} {Phys. Rev. Lett.}\ }\textbf {\bibinfo
  {volume} {107}},\ \bibinfo {pages} {243902} (\bibinfo {year}
  {2011})}\BibitemShut {NoStop}%
\bibitem [{\citenamefont {Kimble}\ and\ \citenamefont
  {Mandel}(1976)}]{kimble76}%
  \BibitemOpen
  \bibfield  {author} {\bibinfo {author} {\bibfnamefont {H.}~\bibnamefont
  {Kimble}}\ and\ \bibinfo {author} {\bibfnamefont {L.}~\bibnamefont
  {Mandel}},\ }\href@noop {} {\bibfield  {journal} {\bibinfo  {journal} {Phys.
  Rev. A}\ }\textbf {\bibinfo {volume} {13}},\ \bibinfo {pages} {2123}
  (\bibinfo {year} {1976})}\BibitemShut {NoStop}%
\bibitem [{\citenamefont {Grimm}\ \emph {et~al.}(2000)\citenamefont {Grimm},
  \citenamefont {Weidem{\"u}ller},\ and\ \citenamefont
  {Ovchinnikov}}]{grimm00a}%
  \BibitemOpen
  \bibfield  {author} {\bibinfo {author} {\bibfnamefont {R.}~\bibnamefont
  {Grimm}}, \bibinfo {author} {\bibfnamefont {M.}~\bibnamefont
  {Weidem{\"u}ller}}, \ and\ \bibinfo {author} {\bibfnamefont {Y.~B.}\
  \bibnamefont {Ovchinnikov}},\ }in\ \href@noop {} {\emph {\bibinfo {booktitle}
  {Advances in Atomic, Molecular, and Optical Physics}}},\ Vol.~\bibinfo
  {volume} {42}\ (\bibinfo  {publisher} {Elsevier},\ \bibinfo {year} {2000})\
  pp.\ \bibinfo {pages} {95--170}\BibitemShut {NoStop}%
\bibitem [{\citenamefont {Zhou}\ \emph {et~al.}(2010)\citenamefont {Zhou},
  \citenamefont {Xu}, \citenamefont {Chen},\ and\ \citenamefont
  {Chen}}]{zhou10}%
  \BibitemOpen
  \bibfield  {author} {\bibinfo {author} {\bibfnamefont {X.}~\bibnamefont
  {Zhou}}, \bibinfo {author} {\bibfnamefont {X.}~\bibnamefont {Xu}}, \bibinfo
  {author} {\bibfnamefont {X.}~\bibnamefont {Chen}}, \ and\ \bibinfo {author}
  {\bibfnamefont {J.}~\bibnamefont {Chen}},\ }\href@noop {} {\bibfield
  {journal} {\bibinfo  {journal} {Phys. Rev. A}\ }\textbf {\bibinfo {volume}
  {81}},\ \bibinfo {pages} {012115} (\bibinfo {year} {2010})}\BibitemShut
  {NoStop}%
\bibitem [{\citenamefont {Shao}\ \emph {et~al.}(2017)\citenamefont {Shao},
  \citenamefont {Huang}, \citenamefont {Guan}, \citenamefont {Li},
  \citenamefont {Shi},\ and\ \citenamefont {Gao}}]{shao17}%
  \BibitemOpen
  \bibfield  {author} {\bibinfo {author} {\bibfnamefont {H.}~\bibnamefont
  {Shao}}, \bibinfo {author} {\bibfnamefont {Y.}~\bibnamefont {Huang}},
  \bibinfo {author} {\bibfnamefont {H.}~\bibnamefont {Guan}}, \bibinfo {author}
  {\bibfnamefont {C.}~\bibnamefont {Li}}, \bibinfo {author} {\bibfnamefont
  {T.}~\bibnamefont {Shi}}, \ and\ \bibinfo {author} {\bibfnamefont
  {K.}~\bibnamefont {Gao}},\ }\href@noop {} {\bibfield  {journal} {\bibinfo
  {journal} {Phys. Rev. A}\ }\textbf {\bibinfo {volume} {95}},\ \bibinfo
  {pages} {053415} (\bibinfo {year} {2017})}\BibitemShut {NoStop}%
\end{thebibliography}
\end{document}